\documentclass[12pt]{iopart}

\usepackage{lipsum}
\usepackage[margin=1in,left=1.5in,includefoot]{geometry}
\usepackage[hidelinks]{hyperref} 

\usepackage{subfig}

\usepackage{setspace}
\usepackage{graphicx} 		 
\usepackage{float}    		 
\graphicspath{{images/}} 

\usepackage[table,xcdraw]{xcolor}

\usepackage{fancyhdr}
\expandafter\let\csname equation*\endcsname\relax
\expandafter\let\csname endequation*\endcsname\relax
\usepackage{amsmath,bm}
\usepackage{siunitx}
\usepackage{tabularx}
\usepackage{caption}

\def\BB{{\bf B}}
\def\DD{{\bf D}}
\def\ee{{\bf e}}
\def\ff{{\bf f}}
\def\FF{{\bf F}}

\def\qq{{\bf q}}
\def\pp{{\bf p}}

\def\rr{{\bf r}}
\def\RR{{\bf R}}

\def\uu{{\bf u}}
\def\UU{{\bf U}}
\def\vv{{\bf v}}

\def\ZZ{{\bf Z}}

\def\QE{\textsc{Quantum espresso}}

\newcommand{\Lagr}{L}

\definecolor{darkgreen}{RGB}{0,160,48}
\definecolor{darkred}{RGB}{160,0,0}

\begin{document}

\title{Reciprocal space temperature-dependent phonons method from ab-initio dynamics}
\author{Ibrahim Buba Garba$^{1,2}$,
Tommaso Morresi$^{1,3}$,
Charles Bouillaguet$^{4}$,
Michele Casula$^{1}$,
Lorenzo Paulatto$^{1}$}
\address{$^1$Institut de Minéralogie, de Physique des Matériaux et de Cosmochimie (IMPMC), Sorbonne Université, CNRS UMR 7590, MNHN, F-75005 Paris, France}
\address{$^2$Physics Department, Federal University Gashua, Gashua 671106, Nigeria}
\address{$^3$European Centre for Theoretical Studies in Nuclear Physics and Related Areas (ECT*), Bruno Kessler Foundation, Trento, Italy}
\address{$^4$Sorbonne Université, CNRS, LIP6, F-75005 Paris, France, 75252 Paris, France}

\begin{abstract}
We present a robust reciprocal-space implementation of the temperature-dependent effective potential method, our implementation  can scale easily to large cell and long sampling time. It is interoperable with standard ab-initio molecular dynamics and with Langevin dynamics. We prove that both sampling methods can be efficient and accurate if a thermostat is used to control temperature and dynamics parameters are used to optimise the sampling efficiency.
By way of example, we apply it to study anharmonic phonon renormalization in weakly and strongly anharmonic materials, reproducing the temperature effect on phonon frequencies, crossing of phase transition, and stabilization of high-temperature phases.
\end{abstract}

\tableofcontents

\section{Introduction}
The most commonly used methods for computing phonon spectra are harmonic and quasiharmonic approximations, in which the potential energy is expanded in Taylor series up to second order in powers of atomic displacements about their equilibrium positions.  However, the small displacement assumption becomes invalid at high temperatures, especially near a phase transition, or even at low temperatures if the energy profile is shallow, or if the quantum nature of nuclei cannot be disregarded, which warrants the need for anharmonic methods. Furthermore, the perturbative treatment of anharmonicity can be problematic in the presence of imaginary phonons, i.e. harmonic approximation has no suitable ground state upon which a perturbative expansion can be built.

Although non-perturbative anharmonic lattice models, particularly the self-consistent phonon (SCP) theory, have been developed as early as the 1950s by Born and Hooton~\cite{born1955statistische,hooton1958_II} and others~\cite{boccara1965scp,koehler1966scp}, these models have originally been applied to rare gases, like Ne, with simple interaction potentials based on \textit{ad hoc} parameters, thereby limiting their application and predictive power.
Over the last two decades, there has been increasing interest in anharmonic methods based on density functional theory (DFT). 
While these methods are diverse in their formulation and implementation, they mainly differ on (i) the method of computing the forces acting on ions, and (ii) how the potential energy surface (PES) is sampled. 
\emph{Ab initio} molecular dynamics (AIMD) simulations sample the potential energy surface including full anharmonicity. In the latter framework, dynamical and transport properties can be computed as time averages over trajectories using either normal-mode analysis (NMA)~\cite{mcgaughey2014vac, mcgaughey2009kl4methods} or velocity autocorrelation function (VACF) method. NMA and VACF have the advantage that both renormalized phonon frequencies and phonon lifetime are obtained from AIMD non-perturbatively and without explicitly computing higher-order force constants. 
The main disadvantage of these methods is the long simulated time required to reach ergodicity and/or targeted frequency resolution.
In the case particle exchange is relevant, an additional drawback is
the lack of well-defined quantum statistics.

Nevertheless, the efficiency of VACF can be substantially improved by projecting the atomic velocities (or displacements) of atoms in a supercell on harmonic phonons modes from DFT in a commensurate unit cell~\cite{dekoker2009VACF, renata2014VACF}.

The early contribution in the family of self-consistent phonon (SCP) methods was the self-consistent \emph{ab initio} lattice dynamics (SCAILD) by Souvatzis~\emph{et al.}~\cite{SCAILD2008entropy}, which is based on thermal mean square displacements of atoms in a supercell. Roekeghem~\emph{et al.}~\cite{QSCAILD2016, QSCAILD_II_2021mingo} extended SCAILD to QSCAILD, which uses a quantum mean
square thermal displacement matrix. The temperature-dependent effective potential (TDEP) approach involves fitting force constants to \emph{ab initio} DFT forces sampled along MD trajectories~\cite{esfarjani2008method, TDEP2011hellman} or stochastically~\cite{TDEP2018stochastic}. Other SCP-based methods include SCPH proposed by Tadano \emph{et al.}~\cite{Tadano_2015}, where anharmonic frequencies are computed from the pole of the Green’s function and higher order effective force constants from ``compressive sensing''~\cite{csld2019zhou}. The
stochastic self-consistent harmonic approximation (SSCHA)~\cite{SSCHA2014errea} minimizes the free energy of a system within a harmonic density matrix ansatz, rigorously capturing both nuclear quantum effects (NQE) and anharmonicity.
Anharmonic effects can also be sampled using a classical Langevin dynamics based on the algorithm by Bussi and Parrinello~\cite{Bussi_Parinello_2007LD} and a recent path integral LD (PILD) scheme using classical and quantum correlators~\cite{pioud_2021, pioud2022_moressi}. 
In PILD, including 
NQE implies a significant expense due to the requirement to simulate many replicas of the system in parallel.
Details of these methods are in the original articles of the authors and also in recent reviews by Esfarjani \emph{et al.}~\cite{esfarjani2018review}, and Hong \emph{et al.}~\cite{hong2021review}.

Here we propose a method that allows efficient prediction of the temperature dependence of phonons and force constants based on the TDEP technique, using a highly efficient reciprocal-space representation that fully exploits  crystal and $\qq$-points symmetries in the Brillouin zone to minimize the number of degrees of freedom.
The rest of the article is structured as follows: In Section 2, we review the TDEP method starting with harmonic approximation, \emph{ab initio} molecular dynamics and Langevin dynamics. Section 3 presents our implementation of the reciprocal space TDEP method, the temperature-dependent phonon (TDPH) method, including its features and some convergence tests.  In Section 4, we apply the TDPH method to treat anharmonicity beyond QHA in weakly anharmonic fcc aluminum and high-temperature $\beta$-phase of Zr. We also study phonon renormalization in SrTiO$_3$ and show how the TDPH method successfully captures temperature-driven hardening of antiferrodistortive mode.

\section{Method}
\subsection{Harmonic Approximation}
The Born-Oppenheimer  potential energy of interacting atoms in a crystal can be expressed as a Taylor series expansion around the equilibrium position:
\begin{equation}\label{eq:Taylor_exp}
V = V_0 + \sum_{i, \alpha}\Phi_{i}^{\alpha} u_{i}^{\alpha} +
\frac{1}{2!}\sum_{ij, \alpha\beta}\Phi_{ij}^{\alpha\beta} u_{i}^{\alpha} u_{j}^{\beta}+
\frac{1}{3!}\sum_{ijk, \alpha\beta\gamma}\Phi_{ijk}^{\alpha\beta\gamma} u_{i}^{\alpha} u_{j}^{\beta} u_{k}^{\gamma} +... \ \ \ .
\end{equation}
Here $\Phi$ are the interatomic force constants (FCs), $ijk$ are atomic indices and $\alpha\beta\gamma$ are Cartesian directions $(x,y,z)$. On the right-hand side, the $V_0$ term is a constant that can be set to zero, while the linear term is zero at equilibrium geometry. The force acting on atom $i$ along 
direction $\alpha$ 
is given as
\begin{equation}\label{eq:Taylor_exp_force}
\begin{aligned}
\FF^{\alpha}_{i} & = -\frac{\partial V}{\partial u^{\alpha}_i}\\ 
& =-\sum_{j, \beta}\Phi_{ij}^{\alpha\beta} u_{j}^{\beta} 
-\frac{1}{2!}\sum_{jk,\beta\gamma}\Phi_{ijk}^{\alpha\beta\gamma} u_{j}^{\beta} u_{k}^{\gamma} +... \  \ \ .
\end{aligned}
\end{equation}

Truncating Taylor's expansion in Equation \ref{eq:Taylor_exp} at the 2nd-order term with respect to the displacement $\uu_{}$, reduces the problem to a well known system of coupled harmonic oscillators that can be solved on a basis of harmonic collective vibrations (phonons) with a wavevector $\qq$, a polarization $\ee_\qq$ and an angular frequency $\omega_\qq$ that solve the following eigenvalue equation:
\begin{equation}\label{eq:eigen-omega}
\DD(\qq) \ee_{\qq, s}= \omega^{2}_{\qq,s} \ee_{\qq, s} \ ,
\end{equation}
where the dynamical matrix $\DD$ is the Fourier transform of real-space FCs $\Phi_{ij}^{\alpha\beta}$, renormalized with the mass $m_i$ of the nuclei :
\begin{equation}\label{eq:dynmat}
\DD_{ij}^{\alpha\beta}(\qq)= \frac{1}{\sqrt{m_im_j}}\sum_{\RR}\Phi_{ij}^{\alpha\beta}e^{-i\qq\cdot\RR} \ .
\end{equation}
The harmonic approximation gives phonons as well-defined quasiparticles with infinite lifetime and independent of temperature.
We note that the Fourier transform is not the only way to build an approximate polynomial for the potential-energy surface, and the global minimum is not necessarily the only possible point around which this expansion can be performed. To have well-defined phonons, the dynamical matrix has to be definite positive, i.e. the second order polynomial approximation has to have positive curvature. More specifically, the Taylor expansion depends on the second derivative of the PES at the equilibrium ionic position. In some cases, the oscillatory motion of the nuclei may cover a relatively large distance and thus see, on average, a very different curvature. This is more common at high temperature, for light ions, and when the PES is close to a transition point, where 
its curvature changes from positive (stable) to negative (unstable).

\subsection{Sampling the Potential Energy Surface (PES)}\label{sec:PES}
We employed two methods of sampling the Born-Oppenheimer PES: the standard molecular dynamics (MD) with Verlet integration of the trajectories, and the Langevin dynamics (LD) based on the integration method by Bussi and Parrinello~\cite{Bussi_Parinello_2007LD, pioud_2021}.
For the MD case, we sampled the PES in two ways, one based on the NVE ensemble (constant number of particles, volume and energy) and the other on NVT ensemble (constant temperature). In the latter case, the stochastic SVR thermostat~\cite{svr2007canonical} is used, which produces a dynamics conceptually halfway between NVE and damped LD.

Generally, the Hamiltonian describing the dynamics of a set of $N$ interacting atoms with coordinates $\rr = (\rr_1, \rr_2, ... \rr_{N})$ and momenta $\pp = (m_1\vv_1, m_2\vv_2, ... m_{N}\vv_{N})$  moving in a potential of the form $V(\rr)$ is 
\begin{equation}
H = \sum_{i} \frac{\pp^{2}_i}{2m_i} + V(\rr) \ .
\end{equation}
$V(\rr)$ could be any analytic potential or computed from first principles electronic structure~\cite{allen_Tildesley2017computer, hutter2000_aimd}. In any case, the aim is to explore the phase space $(\pp,\rr)$ exhaustively so that an ensemble average $\left\langle . \right\rangle_{(\pp,\rr)}$ is equivalent to time average $\left\langle . \right\rangle_{t}$,
assuming ergodicity, for an observable $A(\pp,\rr)$, i.e.
\begin{equation}\label{eq:observable}
\left\langle A \right\rangle = \lim\limits_{t\rightarrow \infty}\int_{0}^{t} A(t')dt' \approx
\frac{1}{N_{steps}}\sum_{i=1}^{N_{steps}} A(\pp(t_i), \rr(t_i)) \ .
\end{equation}
\textit{Ab initio} MD implies solving the static electronic structure problem and simultaneously propagating the nuclei classically, according to the equations of motion
\begin{equation}\label{eq:BOMD}
\begin{cases}
\dot{\pp}_i(t) = \ff_{i} = -\nabla_i V(\rr) \\
\dot{\rr}_i(t) = \frac{\pp_i(t)}{m_i}
\end{cases}\,.
\end{equation}  
The potential energy $V(\rr)$ depends, either parametrically or from the explicit \emph{ab initio} solution of the electronic problem, on atomic positions  $\{\rr_i\}_{i=1}^{N}$. The right-hand side of Equation \ref{eq:BOMD} gives the forces $\ff_i$ acting on each atom $i \in \{1,2,...N \}$. \\
In the case of Langevin dynamics, the nuclei are propagated according to
\begin{equation}\label{eq:pioud}
\begin{cases}
\dot{\pp}_i(t) =- \underbrace{\bm{\gamma}\pp_i(t)}_{\text{damping}} \ \ + \underbrace{\ff_i(\rr(t))}_{\text{Hellmann-Feymann}}
+ \ \underbrace{\sqrt{2m_ik_BT}\bm{\eta}_i(t)}_{\text{stochastic}} \\
\dot{\rr}_i(t) = \frac{\pp_i(t)}{m_i}
\end{cases}\,.
\end{equation}
In the above Equation, $\bm{\gamma}$ is the Langevin friction, $\ff(\qq(t))$ the deterministic force, and $\bm{\eta}(t)$ the white-noise term. Following the fluctuation-dissipation theorem (FDT)~\cite{kubo1966fluct_dis}, the dissipative Langevin damping $\bm{\gamma}$ is compensated by fluctuations induced by the stochastic forces $\bm{\eta}(t)$ via temperature dependence, such that
\begin{equation}
\langle \bm{\eta}(t) \bm{\eta}(t')\rangle = \delta(t-t')\bm{\alpha}(\tau) \text{~~~~with $\bm{\alpha} = 2k_{B}T\bm{\gamma}$.}
\end{equation}
The equations of motion 
in Equation \ref{eq:pioud} can be expressed in terms of the probability density $P(\mathnormal{\Gamma},t)$ that evolves according to the Fokker-Planck equation~\cite{Risken_Fokker_Plank1996} which reads as
\begin{equation}
\frac{\partial P(\Gamma,t)}{\partial t}=i\Lagr_{FP}P(\Gamma,t),
\end{equation}
where $i\Lagr_{FP} = i\Lagr_{p} + i\Lagr_{q} + i\Lagr_{\gamma}$ is the Fokker-Planck Liouvillian, evolving the momenta, the coordinates and defining the thermostat, respectively.\\
There are different 
methods one can choose to integrate the LD of Equation~\ref{eq:pioud}. We have chosen a recent implementation of the Bussi and Parrinello algorithm.~\cite{Bussi_Parinello_2007LD, pioud_2021} 
The major advantages of this method over MD are the small correlation time and fast equilibration, which significantly reduce computing time and allow efficient canonical sampling of the PES.

Originally developed to describe the behavior of a molecule in water, by treating implicitly the interaction with the solvent as damping and random external forces, it has long been shown that LD can properly describe the thermodynamical behavior of more diverse systems~\cite{Bussi_Parinello_2007LD}. In its limiting cases, it reduces to NVE (no damping) or to a Brownian motion (no Born-Oppenheimer forces), while in the intermediate range its behavior depends on the ratio of the damping and the time step, with a large range of acceptable good values which can be determined by testing and computing the correlation time of $V$. 
We found that the value $\gamma = 0.00146 \ a. u.$ is sufficiently close to the optimum for all our applications.
\subsection{Temperature-dependent Effective Potential (TDEP)}\label{subsec_tdep}
Temperature-dependent effective potential (TDEP) is a method of computing finite-temperature force constants (FCs) from a sampling of \emph{ab initio} forces. This method was first proposed by Esfarjani and Stokes~\cite{esfarjani2008method}, where the FCs (harmonic, cubic, and quartic) of Si were extracted from \emph{ab initio} MD force-displacement data. Hellman and coworkers~\cite{TDEP2011hellman} introduced a similar procedure, but starting instead with zero-temperature harmonic FCs to obtain the best effective FCs at a given temperature by a least-square fit to \emph{ab initio} MD forces at that temperature.\\

The TDEP method was tested on entropy-stabilized bcc phases of Zr and Li by Hellman \emph{et al.}~\cite{TDEP2011hellman}, and then later extended to treat anharmonic free energy correction in \textsuperscript{4}He~\cite{TDEP2013free-energy} and cubic FCs in Si and FeSi~\cite{TDEP2013_IFC}. The TDEP formalism starts with a model Hamiltonian
\begin{equation}\label{eq:tdep_harm}
H = U_0 + \sum_{i} \frac{\pp^2_i}{2m_i} + \frac{1}{2}\sum_{ij \alpha\beta}\phi^{\alpha\beta}_{ij}u^{\alpha}_iu^{\beta}_j,
\end{equation}
where $U_0$ is the ground state energy, and $\uu$ and $\pp$ denote displacement and momentum, respectively, of atom $i$ ($j$) in Cartesian direction $\alpha$ ($\beta$). 
Based on the Hamiltonian above, one can define a harmonic approximation of the force $\FF^{H} = -\nabla V$ as
\begin{equation}\label{eq:tdep_f}
\FF^{H} \equiv \FF^{\alpha}_i =-\sum_{j\beta}\phi^{\alpha\beta}_{ij}u^{\beta}_j.
\end{equation}
The model Hamiltonian in Equation \ref{eq:tdep_harm} is then fitted to a Born-Oppenheimer energy surface, sampled at finite temperature, by minimizing the residual mean squared difference between $\FF^{H}$ and the \emph{ab initio} (Hellman-Feynman) force, $\FF^{AI}$, at each time step $t$:
\begin{equation}\label{eq:tdep_ch}
\begin{aligned}
\chi^2 = & \ \frac{1}{N_{step}}\sum_{t=1}^{N_{step}}|\FF^{AI}_t - \FF^{H}_t|^2 \\
= & \ \frac{1}{N_{step}}\sum_{t=1}^{N_{step}}|\FF^{AI}_t - \Theta \UU|^2. \\
\end{aligned}
\end{equation}
The least-square determination of $\FF^{H}$ is then provided by the pseudoinverse solution that gives the lowest residual force:
\begin{equation*}
\Theta = \UU^{\dagger}\FF^{AI} = 
\begin{pmatrix}
\uu_1  \\
\\
:\\
\\
\uu_N
\end{pmatrix}^{\dagger}
\begin{pmatrix}
\FF^{AI}_1  \\
\\
:\\
\\
\FF^{AI}_N
\end{pmatrix}.
\end{equation*}
The fitted FCs, $\Theta = \phi_{ij}^{\alpha\beta}(T)$ can then be used to compute temperature-dependent properties. Due to the computational cost of AIMD, stochastically initialized temperature-dependent effective potential (s-TDEP)~\cite{TDEP2018stochastic} was proposed in which atoms in a supercell are displaced with a stochastic thermal displacement~\cite{Estreicher2006stochastic, SSCHA2014errea} to allow  sampling 
the canonical ensemble. While s-TDEP enables the inclusion of both anharmonic and quantum effects, its stochastic nature results in phase space that is not necessarly consistent with the requested temperature. 
In this work, we adopted the Langevin
dynamics based on the algorithm by Bussi and Parrinello~\cite{Bussi_Parinello_2007LD, pioud_2021}, where instead of using an acceptance probability, as in Monte Carlo algorithms, the additional knowledge of deterministic forces is used to construct a chain of dynamic steps that allows for optimized sampling efficiency. \\

In the case of polar materials, the long-range contribution to FCs due to dipole-dipole interactions, $\Phi^{LR}_{ij}$, are computed using \textit{ab initio} Born effective charges $\ZZ^*$, and dielectric constant $\epsilon^{\infty}$ from density functional perturbation theory (DFPT). Long-range effects are impossible to compute unless one uses a very large supercell. However, it is possible to separate forces into short- and long-range contributions  $\FF^{AI} = \FF^{SR} + \FF^{LR}$, and then fit only the rapidly decaying short-range forces $\FF^{SR}$. Detail explanation will be provided in Section \ref{Polar Materails}. 
\section{Implementation of Reciprocal Space TDEP Method (TDPH)}\label{sec:implementation}
The temperature-dependent phonon (TDPH) method is a reciprocal space implementation of the TDEP technique described in Section \ref{subsec_tdep}. 
Our method is similar to the TDEP technique described above.
However, instead of fitting the FCs directly to \textit{ab initio} forces, we decompose the dynamical matrices $D(\qq)$ on symmetrized basis $\BB_i(\qq)$ at each $\qq$-point~\cite{maradudin1962scattering}. This procedure is inspired to the one used in SSCHA\cite{SSCHA2014errea} and ensures that an irreducible set of parameters is employed to describe the full phonon dispersion. These minimum phonon parameters
(MPP) can be used to recompose the FCs and thus be fitted to the \textit{ab initio} forces.

We will see in Sec.~\ref{symm_basis}
how the basis $\BB_i(\qq)$ is built. Here, we detail the fitting procedure, which can be summarized in the following
steps:
\begin{enumerate}
	\item Compute harmonic phonons on a $\qq$-grid
 and decompose the dynamical matrices $D(\qq)$ on symmetrized basis $\BB_i(\qq)$ at each $\qq$-point
 \begin{equation*}
 	D(\qq) = \sum_{i}^{N_{\BB}}c_i(\qq)\BB_i(\qq) \ ,
 \end{equation*}
  \begin{equation*}
 c_i(\qq)=\left<D(\qq)|\BB_i(\qq)\right> = \mathrm{Tr} [D(\qq) BB_i(\qq)] \ .
 \end{equation*}
    \item Read N configurations of AIMD force-displacement data in a supercell: 
    $\{\FF^{AI}_i,\uu_i\}_{i=1,\ldots,N}$.
    The supercell size must be commensurate with the phonon $\qq$-grid.
    \item If polar material, remove the long-range forces $\FF^{LR}$ (due to dipole-dipole interaction) from the AIMD forces before fit. $\FF^{LR }= \sum_{uc}\sum_{j\beta}\phi^{\alpha\beta(LR)}_{ij}\uu^{\beta}_j$. $\phi^{\alpha\beta(LR)}_{ij}$ is the long-range contribution to FCs due to dipole-dipole interaction defined based on Born effective charges $\ZZ^*$ and dielectric constants $\epsilon^{\infty}$ using DFPT (see Sec.~\ref{polar_materials}).
    \item Minimize the residual force by LMDIF (Modified Levenberg-Marquardt) method
    $$ \label{eq:chi2}
    \chi^2 = \texttt{min}
    \bigg\{\frac{1}{N}\sum_{i=1}^{N}|\FF^{AI}_i - \FF^H_i|^2\bigg\}
    $$
    \item\label{step5} 
    Reconstruct the dynamical matrices from the minimal parameters $c_i(\qq)$.
\end{enumerate}

\subsection{Symmetric Basis for Dynamical Matrices}
\label{symm_basis}
The phonon dispersion over a grid of $n \times n \times n$ \qq-points is described by a $3N_{at} \times 3N_{at}$ Hermitian dynamical matrix at each \qq-point in the grid. In order to reduce the dimension of this space from $n^3 \times N_{at}^2$ down to something more manageable, we can use symmetries in two ways. First, we can reduce the grid of \qq-point to its irreducible wedge, i.e. from every set of \qq-points which are linked by a symmetry operation of the crystal, namely a "star of q-points", we only take one. Second, we observe that at each \qq-point 
the dynamical matrix 
is fully determined by the subset of symmetry operations that leave the \qq-point unchanged, minus a reciprocal space vector~\cite{GPP}.

For every \qq-point in the irreducible wedge, we apply a symmetrization and orthogonalization procedure that from a trial Hermitian matrix of dimensions $N_{at}^2$, with no specific symmetry, returns a minimal orthonormal basis which has the correct symmetry. The symmetrization and orthogonalization algorithm proceeds as follows:
\begin{enumerate}
	\item Start with an initial guess of Hermitian matrices of dimension $(3N_{at})^2$ for each point in the irreducible list. Possible choices are, random matrices, matrices with a single non-zero element, or matrices constructed from the eigenvectors of the zero-temperature DFPT calculation.
	\item The elements of this trial basis are symmetrized, according to the symmetry of the \qq-point. If any element is zero, they are discarded, the others are normalized.
	\item Reduce the basis with the Gram-Schmidt algorithm, because the initial basis is over-complete; if a zero-norm element appear during the procedure it has to be discarded.
\end{enumerate}
Finally, what will be left is a set of symmetry-compatible matrices 1,2,...,$N_{B}(\qq) $ $\forall \; \qq$, such that any dynamical matrix from a simulation of a real crystal at each $\qq$-point can be decomposed in the following way: 
 \begin{equation}
D(\qq) = \sum_{i=1}^{N_{\BB}(\qq)} \langle D(\qq) \;|\; \BB_i(\qq)\rangle \; \BB_i(\qq) \textrm{~~~~~~~$\forall \; \qq \in$ 1BZ}.
\end{equation}
$\BB_i(\qq)$ are the symmetrized basis, defined as the minimal phonon parameters. 
$N_\BB(\qq)$ is also the number of irreducible representation in the point group of $\qq$.
The total number of such parameters is given by
\begin{equation}
N_{\BB} = \sum_{\qq}N_\BB(\qq)
\end{equation}
In practice, $N_{\BB}$ is significantly smaller than $(3N_{at})^2$ and can be as small as 4 for highly symmetric crystals
such as fcc Al on $2 \times 2 \times 2 \; \qq$-grid. 
Table \ref{Tab:table_params} below gives the phonon parameters of different materials. The number of phonon parameters $N_{\BB}$, depends on the symmetry of the crystal structure and the size of the phonon $\qq$-grid (which determine the size of the supercell in real space). For instance, $N_{\BB}$ of \textit{orthorhombic} MgSiO$_3$ is 19 times bigger than that of \textit{cubic} MgO, although the latter have 1.6 times more atoms than the former.
\begin{center}
\captionof{table}{Size dependence of phonon parameters for a 5,000  MD snapshots. Space group (SG), supercell size, number of atoms in supercell ($N_{at}$), minimal phonon parameters ($N_\BB$), \textit{ab initio} forces for $N_{steps}$ ($F_{AI}$), and $\texttt{TDPH}_{\texttt{time}}$ CPU time (Machine specification:  Personal computer with 2.20GHz (x12) Intel Core i7-8750H ). \label{Tab:table_params}}
		\begin{tabular}{ m{2.cm} m{2.cm} m{2.cm} m{1.cm} m{1.cm} m{2.cm}  m{2.2cm} } 
		\hline
		 System & SG &  Supercell & $N_{at}$ & $N_\BB$ & $F_{AI}$ & $\texttt{TDPH}_{\texttt{time}}(s)$\\ \hline \hline
		Al  &\textit{Fm3m 225}     & ($2 \times 2 \times 2$) & 8 & 4     & 2.00$\ \times 10^{5}$ & 3 \\
		CsI &\textit{Pm3m 221} & ($2 \times 2 \times 2$) & 16 & 13   & 1.20$\ \times 10^{5}$  & 9 \\
		Zr  & \textit{Im3m 229}     & ($4 \times 4 \times 4$ & 64 & 17     & 9.60$\ \times 10^{5}$ & 3 \\
		SrTiO$_3$ &\textit{Pm3m 221} & ($2 \times 2 \times 2$) & 40 & 49   & 6.00$\ \times 10^{5}$  & 42 \\
		MgO &\textit{Fm3m 225}    & ($4 \times 4 \times 4$) & 128 & 51 &1.92 $\times 10^{6}$    & 98\\
		MgSiO$_3$ &\textit{Pbnm 62}  & ($2 \times 1 \times 2$) & 80 & 994  &1.20$\ \times10^{6}$ &  28152\\ 
		\hline
	\end{tabular}
\end{center}
\vspace{.5cm}
\subsection{Dealing with Polar Materials}\label{Polar Materails}
\label{polar_materials}
In polar materials, the long-range nature of the Coulomb forces is responsible of a macroscopic electric field for longitudinal optical phonons (LO), giving rise to the LO-TO splitting as $\qq\rightarrow 0$~\cite{DFPT_2001}. Thus, in polar crystals, the force constants can be separated into an exponentially decaying short-range part $\Phi^{SR}$, and a long-range dipole-dipole term $\Phi^{LR}$~\cite{BornHuang1954, Cowley1962non_anal} that decays polynomially ($\sim d^{-3}$, where $d$ is the interatomic distance):  
\begin{equation}\label{eq:FCs_sr+lr}
	\Phi^{\alpha\beta}_{ij} = 	\Phi^{SR} + \Phi^{LR}.
\end{equation} 
The long-range 
FCs are given by the general form in terms of the Born effective charge tensor $\ZZ_i^{*}$ of the 
$i$-th
atoms in a unit cell and the macroscopic dielectric tensor $\epsilon^{\infty}$ which can be routinely computed within DFPT~\cite{DFPT_2001, gonze1994rigid, baroni1991_long-range} 
\begin{equation}
 \Phi^{LR}\equiv \phi^{\alpha\beta(LR)}_{ij} =\frac{4\pi e^2}{\Omega} \  \frac{(\qq\cdot \ZZ_i^{*})_\alpha(\qq\cdot \ZZ_j^{*})_\beta}{\qq\cdot\epsilon^{\infty}\cdot\qq}
\end{equation}\label{eq:gonze_long_range}
Although the description of the $\Phi^{LR}$ in terms of the dipole-dipole term yields accurate results in the majority of materials, it has been demonstrated that generalizing to higher order multipolar terms is required in some cases~\cite{royo2020multipolar}.
From Equation \ref{eq:FCs_sr+lr}, the atomic forces can also be separated into short-and long-range terms
\begin{equation}
\begin{aligned}
\FF^{AI} = \ & \FF^{SR} + \FF^{LR}\\
         = \ & -\sum_{j\beta}\Phi^{\alpha\beta, SR}_{ij} \uu^{\beta}_j-
           \sum_{j\beta}\Phi^{\alpha\beta, LR}_{ij} \uu^{\beta}_j
\end{aligned}
\end{equation}

Following Ref.~\cite{TDEP2018long-range_sto,csld2019long-range}, 
the short-range FCs, $\ \Phi^{SR}$, are fitted to the short-range forces  $\FF^{SR}$ only. Long-range effects on FCs require a very large supercell, which is not feasible in practice. We tested fitting the effective charges (the long-range forces) and observed weak temperature dependence. This is expected since any change in $\ZZ^*$ would require a considerable change in the electronic structure (which can be treated at the QHA level), but the temperature does not significantly affect the electronic occupations. 
Note that eq.~\ref{eq:gonze_long_range} is in reciprocal space, in order to bring it to real space we compute it for $\qq=0$ in the supercell dual to the $\qq$-grid.
\subsection{Features of TDPH}
The initial guess for FCs could come from DFPT or frozen phonons in a supercell. 
In the case of DFPT,  this implies, among other factors, handling polar materials in one integrated way without the need for a separate or complementary calculation of the long-range dipole-dipole term.
For this reason, no arbitrary real-space cutoff has to be imposed to keep the calculation fast. We have 
verified that after starting with initial symmetric random phonon parameters the fitting procedure converged to the correct result in all of our test cases. In principle, it is possible to start with no initial guess at all (other than the crystal geometry).

As we have described in Sec.~\ref{symm_basis}, the dynamical matrices are decomposed on a set of symmetrized minimal basis at each $\qq$-point in the 1BZ, whose linear coefficients are the parameters that are minimized during the fitting. Thus, this reciprocal-space approach is highly efficient and ensures very fast convergence.

Effective charges can be considered fixed or minimized. We verified that finite temperature effects on the effective charges beyond the QHA level are vanishingly small.

TDPH, as part of the D3Q~\cite{paulatto2013D3Q} code, is fully integrated into the Quantum Espresso (QE) package~\cite{QE-2009,QE_2017}, which computes both FCs and MD configurations. This unique synergy, as well as faster convergence, makes it easier and more efficient to compute anharmonic properties. 
To decorrelate more quickly the sampled configurations,
TDPH can be integrated with both classical and quantum Langevin dynamics, the latter implemented via the PIOUD engine~\cite{pioud_felix_2017} (also based on QE) which is currently still in closed development.

\subsection{Numerical Minimization}
As we have seen in Table~\ref{Tab:table_params}, the number of phonon parameters can be of the order of 1000 for relatively large systems, which may require a few thousand MD steps for the TDPH procedure to accurately converge. 
The Levenberg-Marquardt\cite{more} minimization algorithm needs to solve linear least
squares problems repeatedly.  More precisely, this means finding a vector $x$
that minimizes $\| Ax - b \|$, where $A$ is a rectangular matrix of dimension
$N_p$ (number of parameters) times $N_{data} = 3 \times N_{at} \times N_
{steps}$ (number of Cartesian direction, atoms in the supercell, and of
steps, respectively). The latter can be of the order of one million. Solving
linear least square problems is done by computing a QR factorization of the
matrix --- this is the most computationally expensive step.  This takes hours
using the naive serial algorithm implemented in MINPACK [45]. For this
reason, we parallelized the existing sequential MINPACK package from 1980.
The main modification consists in using ScaLAPACK [46] to compute QR
factorizations.
The implementation we used, minimizes the $N_{data}$ residual forces simultaneously, using their gradient from numerical differentiation. The algorithm is very robust and can converge, even from 
randomly initialized FCs, in a few (typically less than 10) iterations. 
\subsection{Convergence Tests}
The decay time of autocorrelation is the most important parameter in determining the efficiency of a dynamic algorithm to provide thermodynamic sampling. For instance, if the MD steps remain correlated for a long time, then more steps need to be sampled and this will eventually skyrocket the computational cost. Weakly correlated configurations, on the other hand, imply that less computing time will be sufficient to sample the PES by keeping the same target accuracy. Finite-size effects, MD integrators, and thermostats are other factors worth observing. Therefore, we perform the following convergence tests:
\begin{itemize}
  \item Study the performance of TDPH (i.e. convergence of phonon frequencies, $\chi^2$,for a given set of  minimal phonon parameters $N_\BB$) with sampling time 
  \begin{align}
  \tau_{\texttt{sampling}} = \texttt{n\_skip} \times dt\ ,
  \end{align}
  where $dt$ is the AIMD time step and $\texttt{n\_skip}$ is the sampling interval.
  \item Compare the TDPH results for different supercell sizes.
  \item Study the dependence of the sampling efficiency on $\tau_{\texttt{sampling}}$ in MD versus LD, and the optimal $\tau_{\texttt{sampling}}$ to achieve reasonable convergence for properties of interest in both cases.
 \item Universality of FCs fitted over different dynamics methods.
\end{itemize}

Note that from a numerical point of view, what matters is to have \texttt{n\_skip} as small as possible, i.e. to use the computed DFT data efficiently. This can be achieved by increasing the time step, but if it is too large, other properties of the simulation will suffer: the simulation temperature may go out of control, or a bias could be introduced. We will see these points in detail as we study specific cases.

\subsubsection{Convergence of Phonons with AIMD Sampling}
Taking a simple example as a benchmark, we examine the phonon frequency convergence by sampling 200 MD snapshots over different simulations lengths (corresponding to five different sampling times: $\tau_{\texttt{sampling}}$ = 0.1, 0.2, 0.5, 1.0 ps) at 300K. Details of calculations are provided in Appendix~A. As shown in Figure \ref{fig:F_skip}, the phonon spectra converged at $\tau_{\texttt{sampling}}=$ 0.2ps for a $2 \times 2 \times 2$ fcc Al supercell. 
\begin{figure}[H]
	\subfloat{(a)}
	\centering
\includegraphics[height=2in,width=2.6in]{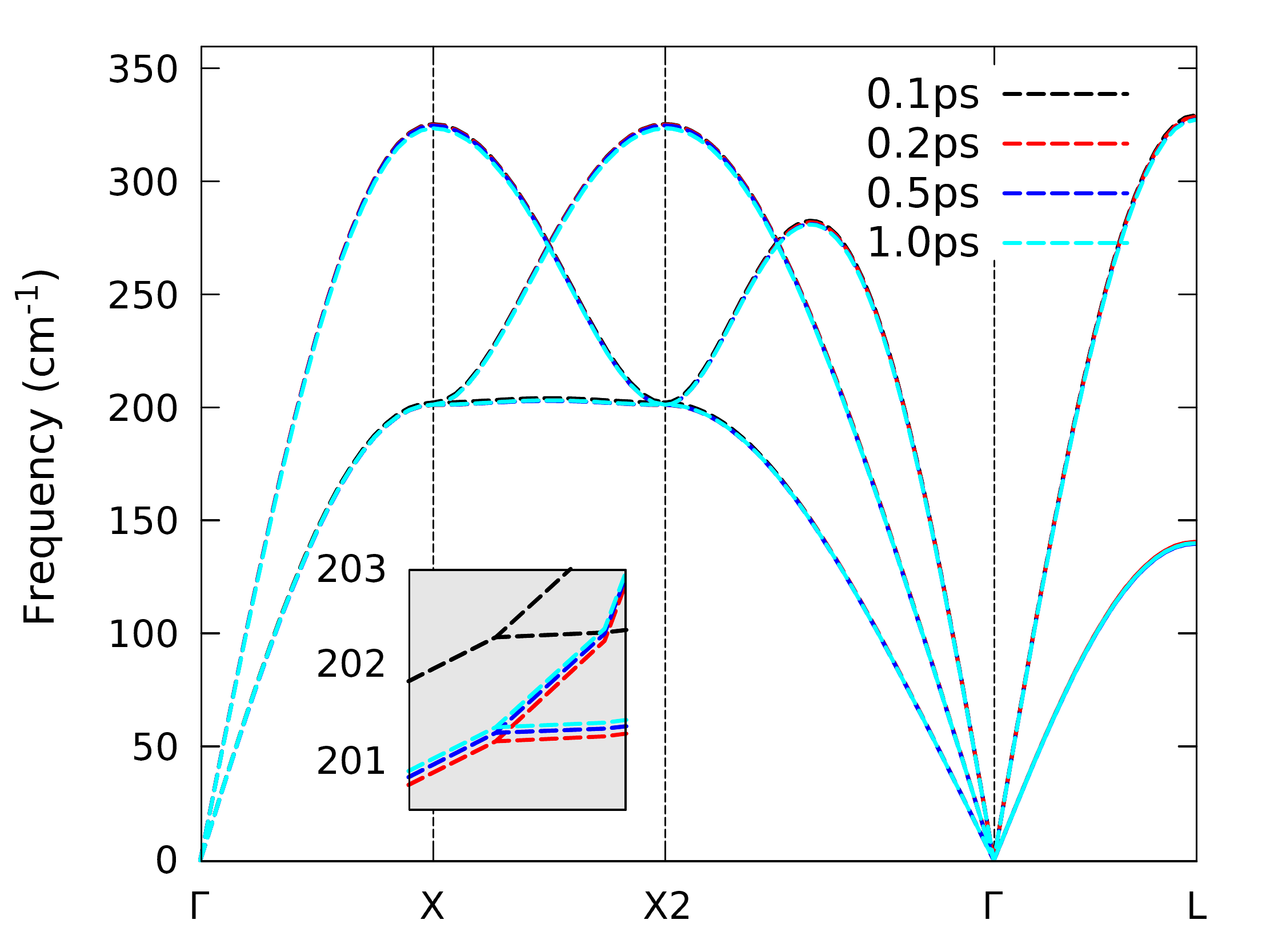}
	\subfloat{(b)}
\includegraphics[height=2in,width=2.6in]{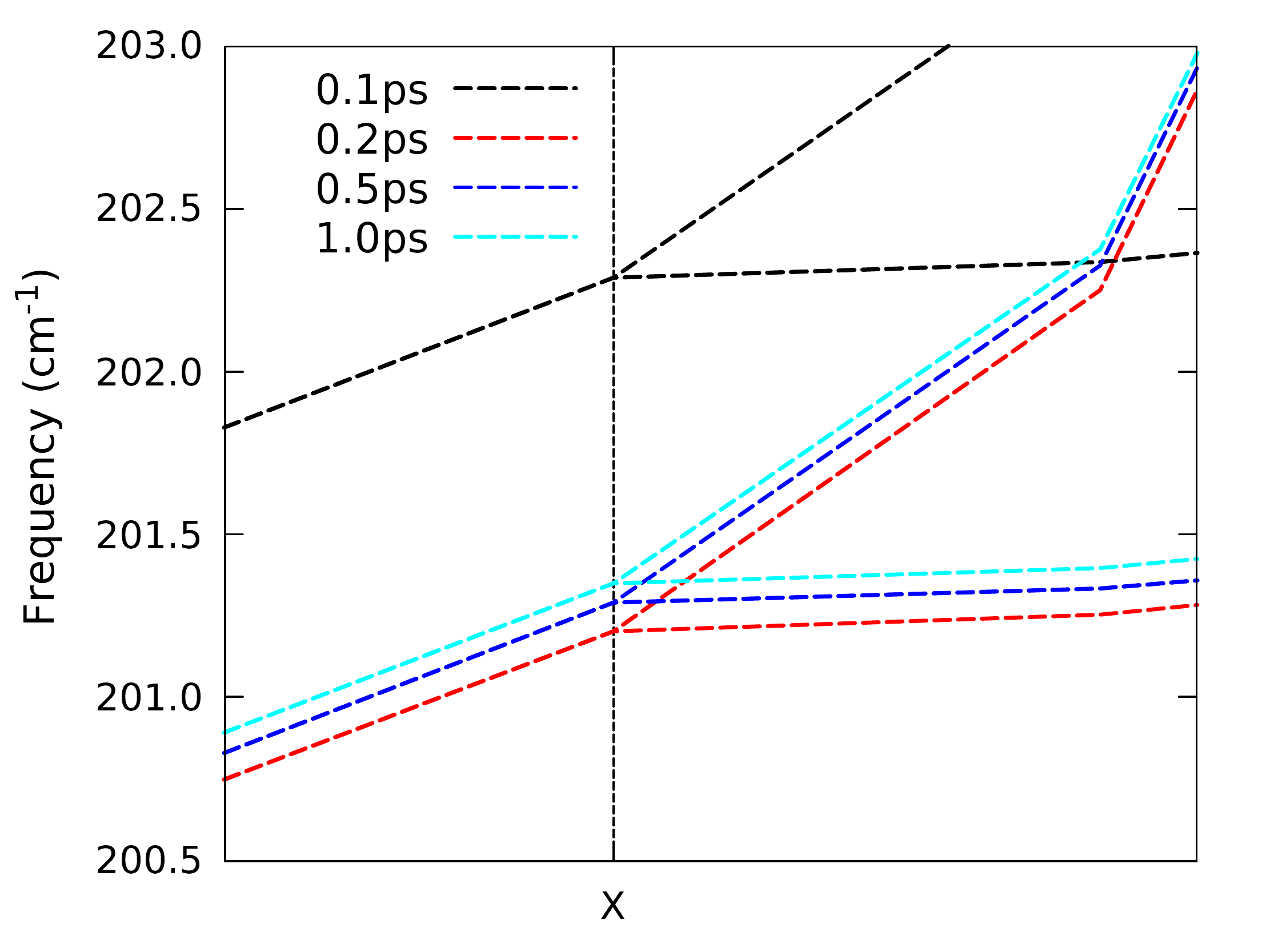}
\caption{Convergence of phonon spectra with respect to the AIMD correlation time. The full phonon dispersion is given in panel (a) and zoomed-out around X in panel} (b).	\label{fig:F_skip} 
\end{figure}

\begin{figure}[H]
	\subfloat{(a)}
	\centering
\includegraphics[height=2in,width=2.5in]{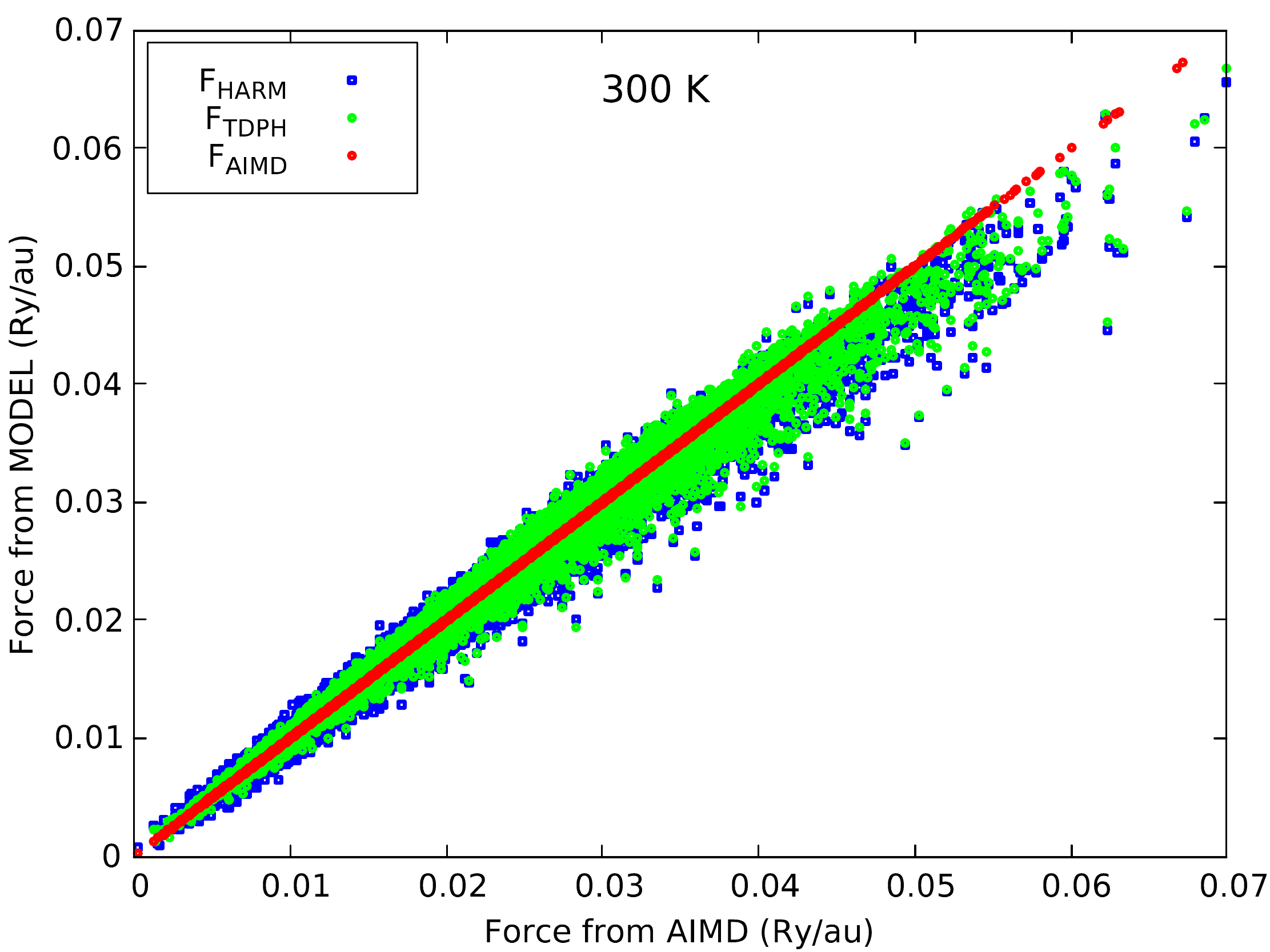}
	\subfloat{(b)}
\includegraphics[height=2in,width=2.5in]{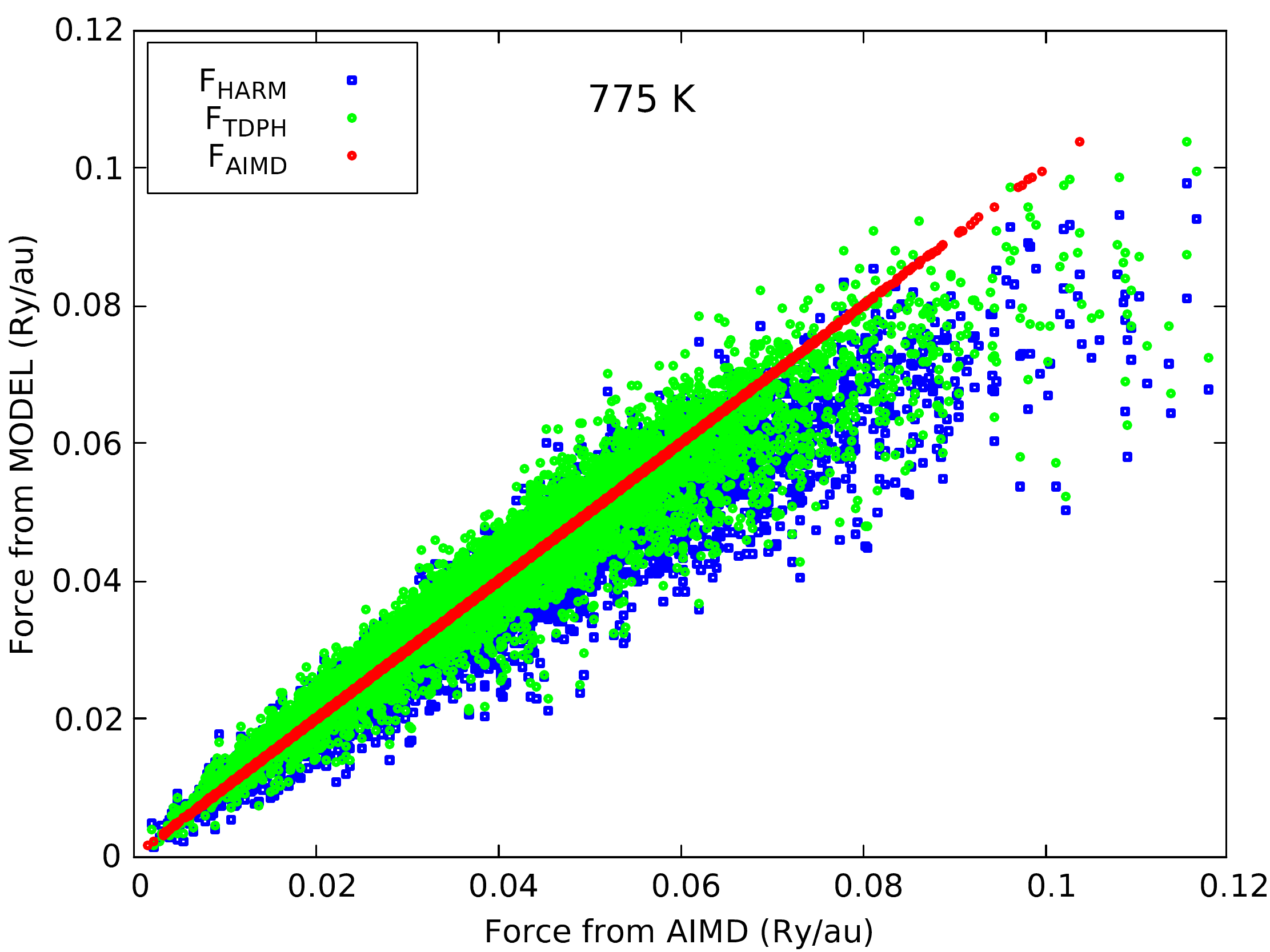}
	\caption{Comparison of the modulus of AIMD force and model (harmonic and TDPH) forces
 at 300 K (panel (a)) 
 and at 775 K (panel (b))}.	\label{fig:F_ratio} 
\end{figure}

\begin{figure}[H]
	\centering
	\includegraphics[height=3in,width=4in]{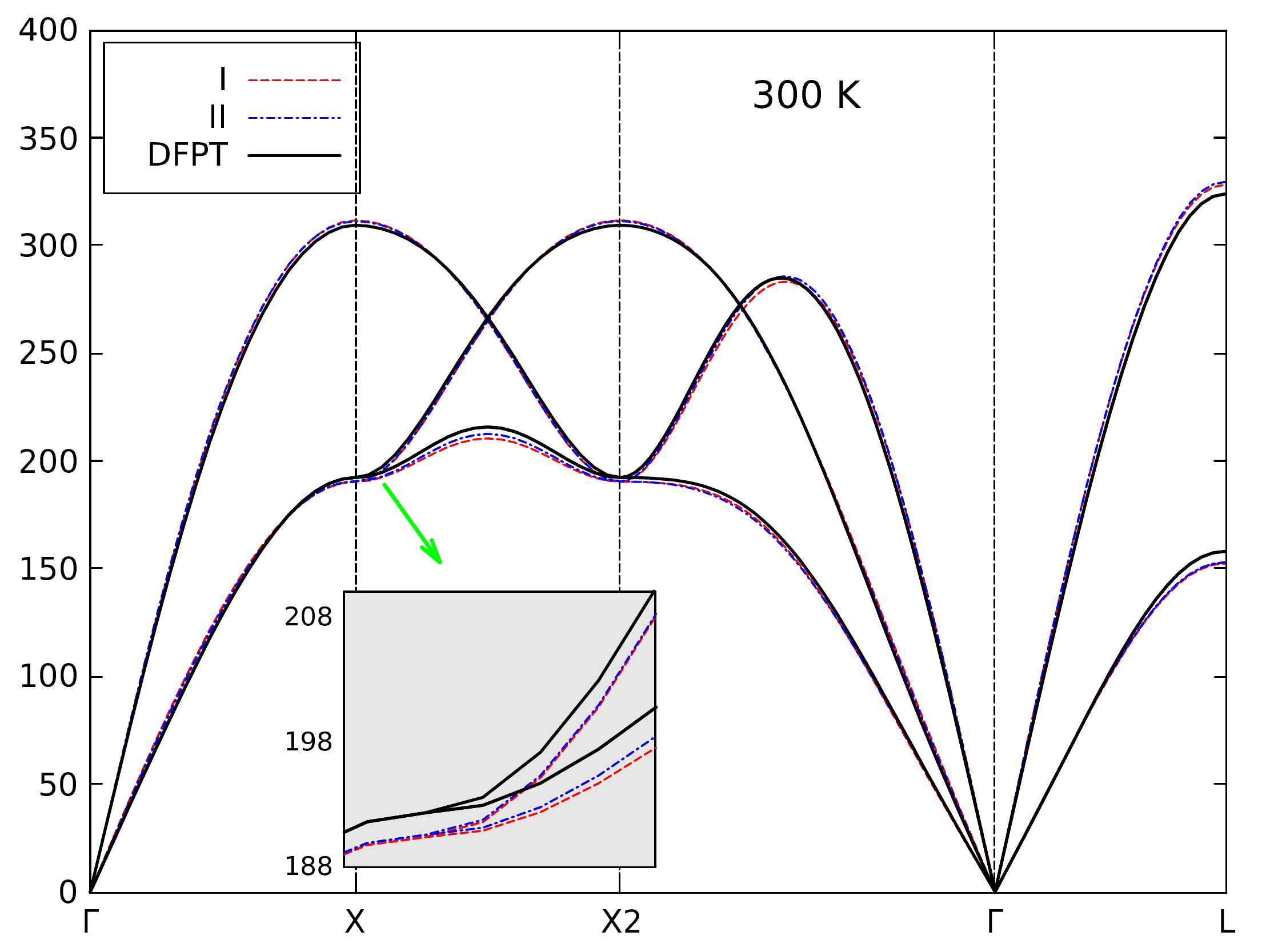}%
	\caption{Comparison of phonons dispersions from two randomly-initialized MD simulations and DFPT for a $3 \times 3 \times 3$ Al supercell.}
\label{fig:runs_vs_phonon_params}  
\end{figure}

As a way to check how TDPH improves the description of atomic forces, we compare the \emph{ab initio} and harmonic forces before and after
fitting the configurations along the MD trajectory. In Figure \ref{fig:F_ratio}, this exercise is done for 400 AIMD snapshots.  
Before the fit, the harmonic forces deviate from the \textit{ab initio} 
ones represented by the dashed circles. After fitting, the TDPH
forces become closer to the \textit{ab initio} determination. Figure \ref{fig:F_ratio} gives the same result at 775K. We observe that the \textit{ab initio} and model forces are closer at a lower temperature (300K), Figure \ref{fig:F_ratio}(a), than at a higher temperature (775 K), Figure \ref{fig:F_ratio}(b). The residual spread around the \emph{ab initio} reference (red line in Fig.~\ref{fig:F_ratio}) gives an idea of the system anharmonicity.
We also examine the robustness of phonon convergence 
with respect to arbitrary initial conditions, by analyzing
different randomly initialized MD runs. Two configurations (system I and II) were initialized with different random atomic velocities, and the TDPH phonons were compared. The result is shown in Figure \ref{fig:runs_vs_phonon_params}, where a nice agreement is found between the two calculations.
\subsubsection{Phonon Parameters versus Supercell Size}\label{sec_NB_vs_sc}
\begin{figure}[H]
	\centering
	\includegraphics[height=3in,width=4in]{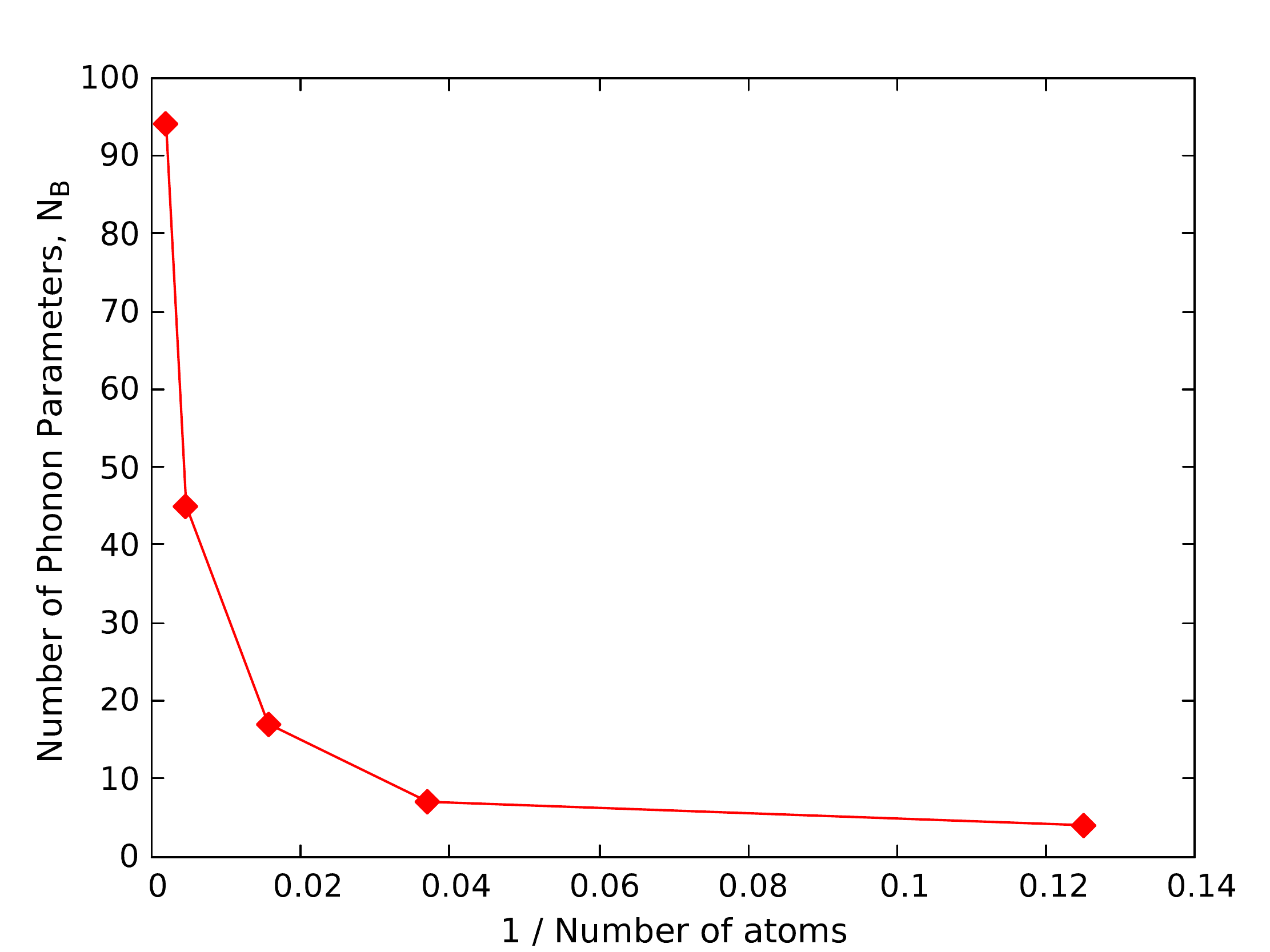}%
\caption{Dependence of the number of phonon parameters $N_{\BB}$ on the number of atoms $N_{at}$ for Al as reported in Table \ref{Tab:table_sc} .}	\label{fig:phonon_params_vs_nat}  
\end{figure}

We test the TDPH performance 
with different supercells 
to examine 
the dependence of the number of phonon parameters $N_{\BB}$ on the supercell size. Different supercells were considered using the same temperature, i.e. $T= 300K$, and same correlation time, namely $\tau_{\texttt{sampling}}=0.2ps$. The result for fcc Al is given in Table \ref{Tab:table_sc}. 
Also shown in Figure \ref{fig:phonon_params_vs_nat}, is the dependence of $N_{\BB}$ on the number of atoms corresponding to different supercell sizes. Note that larger supercells do not necessarily require longer simulations time, since the ratio $N_{F_{AI}}/N_{\BB}$, where $N_{F_{AI}}$ is the number of force components in the supercell, does not change significantly, thanks to the full exploitation of symmetry relations in TDPH.
\begin{center}
	\captionof{table}{Phonon parameters versus supercell size in Al. 
 \label{Tab:table_sc}}
	\begin{tabular}{ m{2cm} m{2cm} m{2cm} m{2cm} m{2cm} }
		\hline \hline
		Supercell & $N_{at}$ &  $N_{\BB}$& $N_{F_{AI}}$ & $N_{F_{AI}}/ N_{\BB}$ \\ \hline \hline
		$2 \times 2 \times 2$ & 8     & 4  & 24   & 6       \\
		$3 \times 3 \times 3$ & 27    & 7  & 81   & 11.6    \\
		$4 \times 4 \times 4$ & 64    & 17 & 192  & 11.3    \\
		$6 \times 6 \times 6$ & 216   & 45 & 684  & 14.4    \\
		$8 \times 8 \times 8$ & 512   & 94 & 1536 & 16.3    \\
		\hline
	\end{tabular}
\end{center}
\subsubsection{Convergence of Phonon Parameters with PES Sampling}
As described in Section \ref{sec:PES}, two schemes were employed to sample the PES in the NVT ensemble, namely MD and LD. For MD, a stochastic velocity rescaling (SVR) thermostat was used, since it has been shown to have no effects on dynamical and transport properties
\cite{svr2007canonical} (see Appendix A for computational details).\\

\begin{figure}[H]
	\centering
	\subfloat{(a)}
	{\includegraphics[scale=0.275]{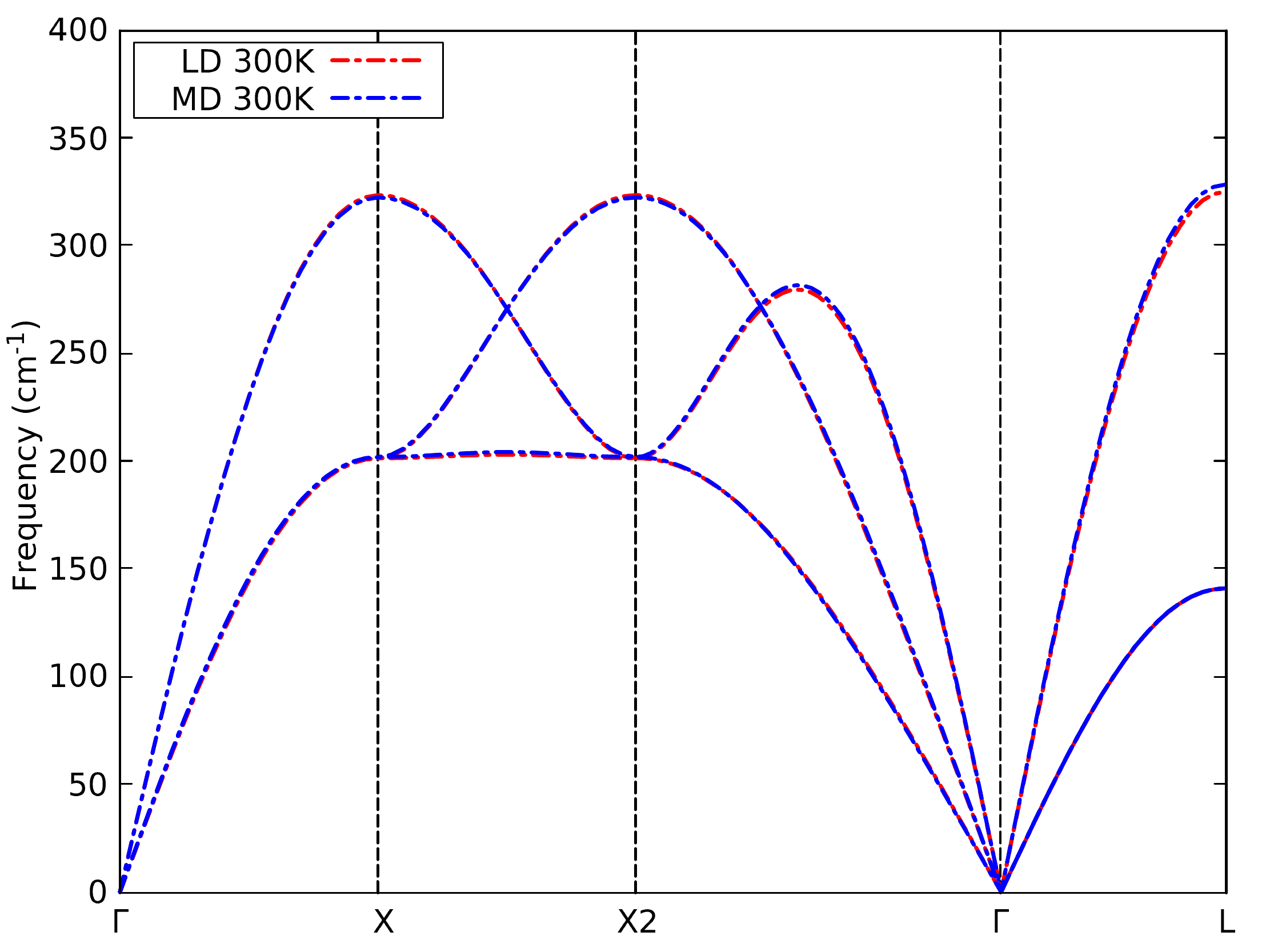}}
	\subfloat{(b)}
	{\includegraphics[scale=0.275]{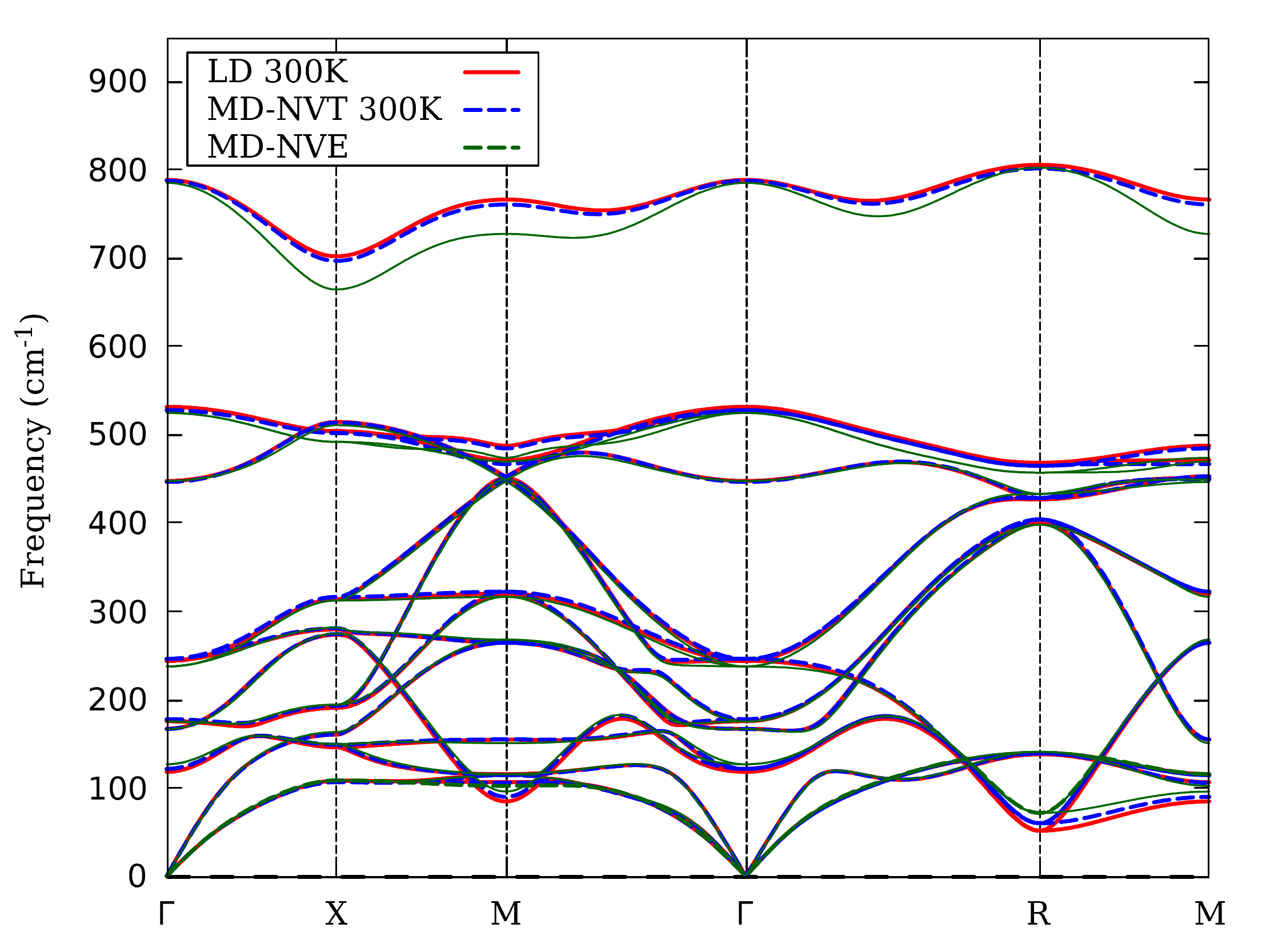}}
\caption{Temperature-dependent phonon dispersion at 300K 
for fcc Al (panel (a)) and 
cubic SrTiO$_3$ (panel (b)) using LD (red line), MD-NVT (blue-dotted line) and MD-NVE (dark-green dotted line). }\label{fig:MD2_vs_LD}
\end{figure}

The second method is the classical Langevin dynamics (LD), which integrates the equations of motion with deterministic (in our case DFT-based) and stochastic forces at a given temperature, stabilized by an appropriate friction, determined though the FDT. It is based on Bussi and Parrinello algorithm~\cite{Bussi_Parinello_2007LD, pioud_2021} using a Trotter factorization of the Liouvillian operator, as already described in Section \ref{sec:PES} and Ref.~\cite{pioud_felix_2017}. Within this formalism, fully anharmonic quantum dynamics can be investigated. By replacing LD with PILD, it is possible to take into account both temperature and 
nuclear quantum effects (NQE),
as recently demonstrated in diamond and high-pressure atomic phase of hydrogen~\cite{pioud_2021}. For our purpose, only thermal effects are considered and the classical LD algorithm is used, since NQE are less significant at higher temperature in the systems studied here. 
As shown in Figure \ref{fig:MD2_vs_LD}(a),
for weakly anharmonic materials like fcc Al, both MD and LD give reasonably converged phonons with 
fewer (i.e. $\sim100$) snapshots and no interval between subsequent steps (i.e. using $\tau_{\texttt{sampling}} = 1$).
In the case of cubic SrTiO$_3$, a strongly anharmonic perovskite, each dynamic simulation gives converged phonons only when an \textit{appropriate} sampling time is used. We choose this sampling time to be the autocorrelation time $C(t,t')$ of the \emph{ab initio} energy (see Appendix C). Using a large time step can considerably reduce $C(t,t')$ (and hence computation time) at the expense of large temperature fluctuations, but this impact is more pronounced in MD. In the case of LD, FDT guarantees that temperature and friction are coupled, keeping the temperature near to its intended value.

For a material with dynamical instability, finite-temperature sampling of the PES with NVE will directly descend to the minima and hence give fast convergence of 
those parameters that correspond to 
soft phonons.
However, sampling the PES using NVT (LD or MD with SVR thermostat) will allow the system to explore the phase space more exhaustively before descending into the minima. They are 
therefore more ergodic than NVE. 
In both cases, the effective harmonic Hamiltonian is constructed to give converged phonon parameters and phonon dispersion representative of the sampling temperature. Figure \ref{fig:MD2_vs_LD}(b) displays the comparison of phonon dispersions yielded by the three different methods. 
\begin{figure}[H]
	\centering
	\includegraphics[scale=0.5]{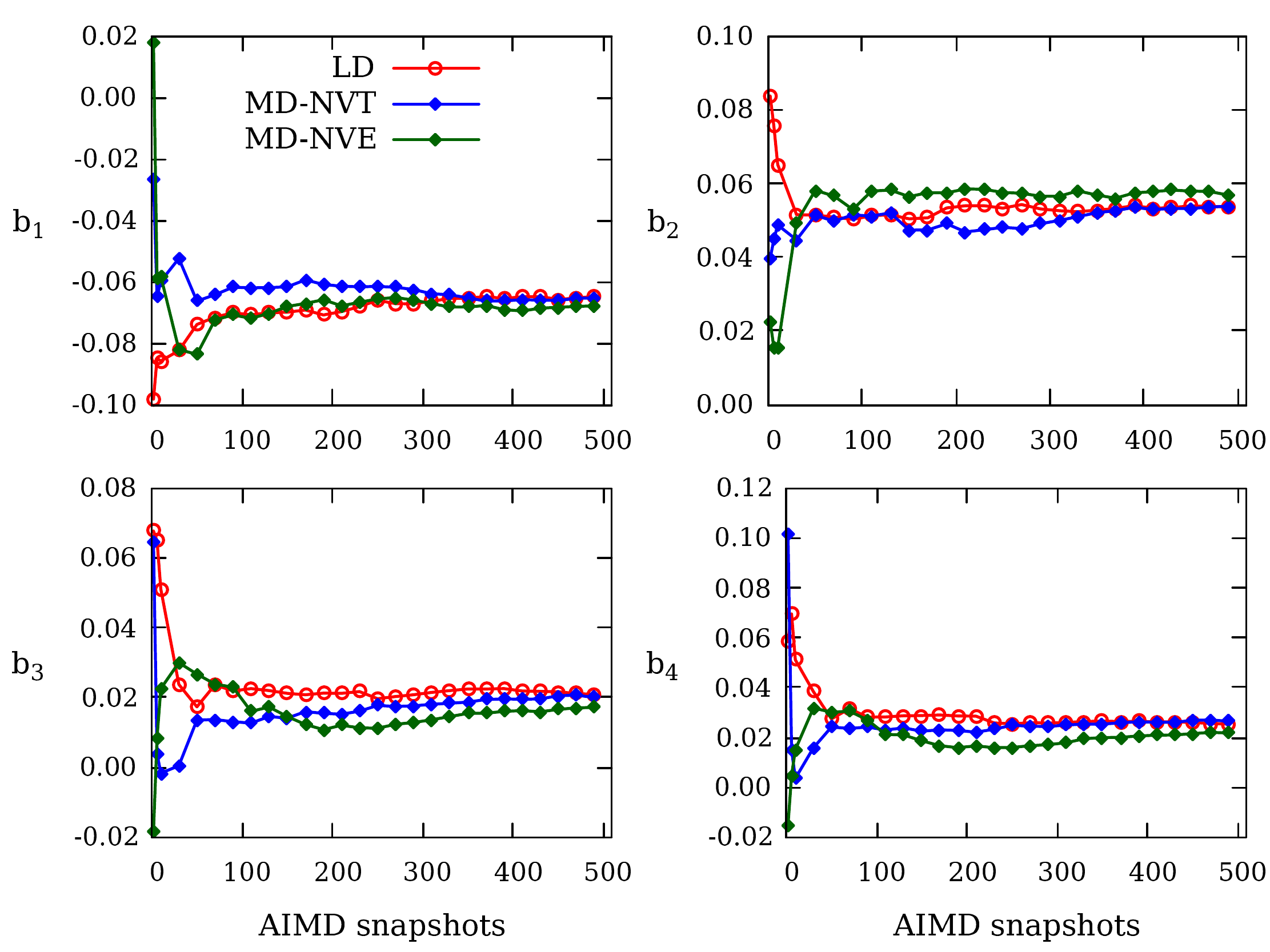}
	\caption{Convergence of the first four phonon parameters in SrTiO$_3$ with TDPH minimization steps. 
All methods are in a reasonable agreement after a 2ps long simulation, with LD yielding the smoothest convergence as a function of the number of AIMD snapshots.}
\label{fig:gr-opt_basis} 
\end{figure}
\subsubsection{Universality of the FCs in different dynamics}
Another factor to take into account is the absence of sampling bias and the ergodicity of the dynamics trajectory i.e. the property of the trajectory to sample the phase space in a way that is consistent with the thermodynamic average. To ascertain this property, we cross-tested FCs from the different dynamics methods (LD, MD- NVT, MD-NVE). To do so, on the one hand we define $\chi^2_{fit}$ of the FCs as the value of $\chi^2$ obtained at the end of the minimization (step~\ref{step5} in the procedure we detailed at the beginning of Sec. ~\ref{sec:implementation}). On the other hand, we define $\chi^2_{test}$ as the $\chi^2$ value
obtained by applying the FC obtained on a given trajectory to a different trajectory, without further minimization. If there is no sampling bias, $\chi^2_{fit}$ and $\chi^2_{test}$ should be very close.

Our assumption is that a good quality trajectory produces universal TDPH FCs, which would work equally well on any other trajectory, i.e. 
$\chi^2_{fit}$ from a good trajectory should not be significantly smaller than that obtained by applying the final FCs to \emph{a different} trajectory.

Conversely, the fact that the $\chi^2$ obtained from a trajectory is too small, does not imply that the trajectory is good. In the best case it indicates that the system is weakly anharmonic, while in the worst case it suggests
that the sampling has a bias. This last point is particularly important: a biased trajectory, which does not explore the entire phase space, may have a smaller $\chi^2$, as the phonon parameters that describe the unexplored degrees of freedom will be unconstrained.

\begin{figure}[H]
\centering
\includegraphics[scale=0.5]{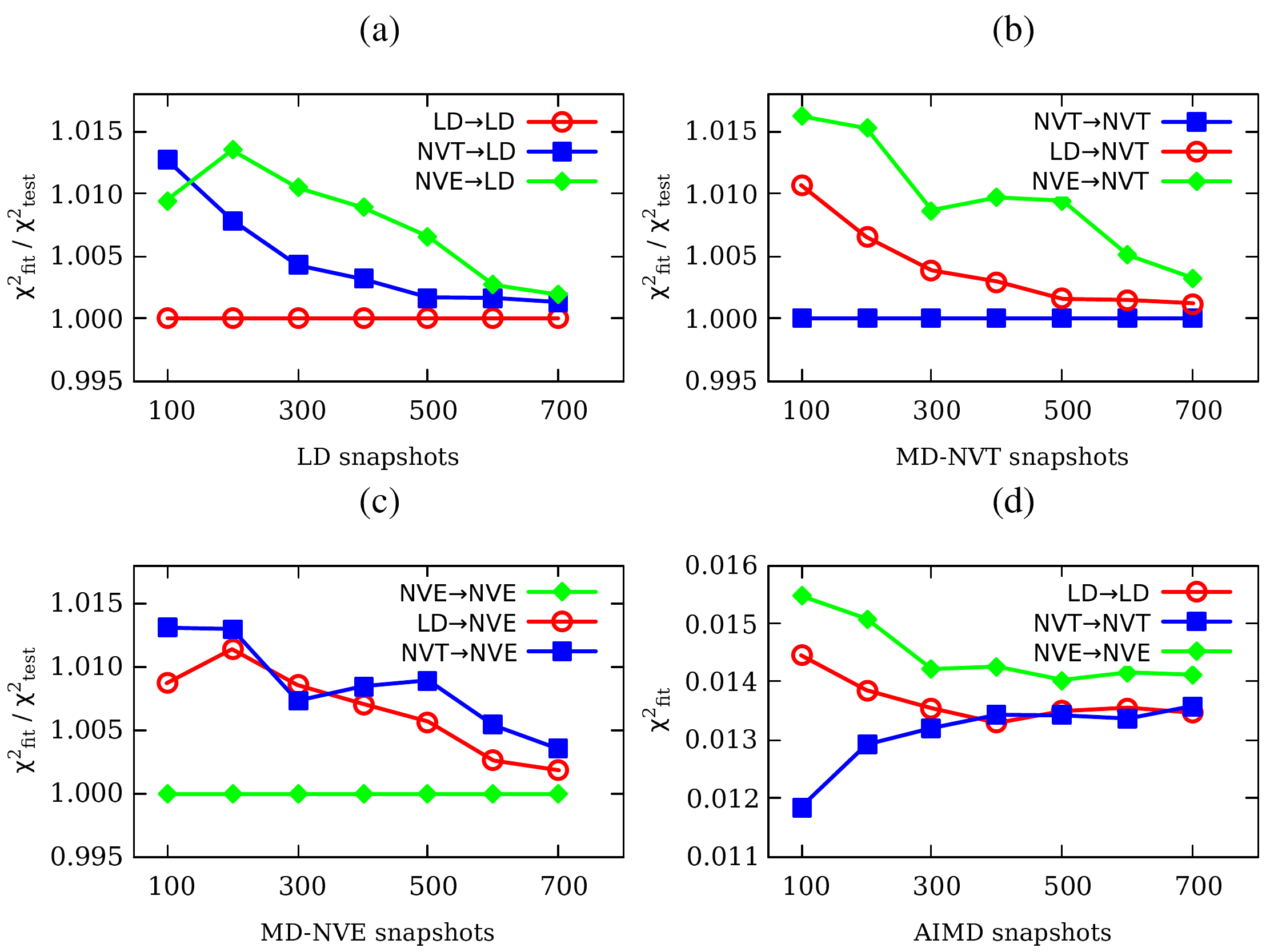}
\caption{Measure of the universality of FCs extracted via TDPH from different dynamics. Horizontal lines 
are obtained by the baseline 
dynamics that generates the ensemble against which each set of FCs is tested. Red, blue and green lines indicate FCs fitted on LD, MD-NVT, and MD-NVE trajectories, respectively.}
\label{fig:gr-cross_all} 
\end{figure}

Figure \ref{fig:gr-cross_all} gives a measure of the universality of FCs in SrTiO$_3$ extracted with the TDPH method from different dynamics: LD, MD-NVT, MD-NVE. In each case, the tested chi-square ($\chi^2_{test}$) is rescaled by the baseline ($\chi^2_{fit}$).
For a sufficiently large number of steps, all three methods seem to give comparable results. We remark however that a tiny difference remains between LD and NVT, which may indicate an equally tiny bias in the exploration of the phase space. On the other hand, there is a larger discrepancy between the uncontrolled NVE and the two other methods. The fact that the NVE $\chi^2$ is higher (panel d of Figure \ref{fig:gr-cross_all}) indicates that it explores a less harmonic region of the phase space.This seems in contradiction with the NVE property of exploring the PES minima more frequently than the other dynamics. In fact,  
we have verified that, for this specific test, the average temperature of the NVE simulation  
drifts from the initial one and settles around $315$~K.
This is because, in a microcanonical ensemble, the total energy drifts due to error accumulation from the MD integrator, which tends to increase with simulation length~\cite{moustafa2018thermostating}. This discrepancy can be particularly problematic when studying the temperature evolution of a system, or when volume expansion is taken into account non-self consistently (i.e. from the equation of state), which is the most common approach.  
Nevertheless,
the temperature drift of NVE can be mitigated by using the average temperature instead of the target one, or by repeating the simulation until the correct temperature is achieved, but this is expensive and labor-intensive. 
Therefore, it turns out that
using a stochastic thermostat can achieve a good result in a much simpler and efficient way.

We can conclude that all methods 
produce universal FCs, but NVT and LD are more accurate and consistent to each other, as the real trajectory temperature is consistent with the desired one ($299$~K for LD and $298$~K for NVT). The $\chi^2$ of the NVE simulation remains consistently higher, which 
is related to the fact that, for this specific test, the effective temperature reached in NVE was larger, leading to the exploration of a more anharmonic region.
\section{Applications}
\subsection{Aluminum: quasiharmonic versus anharmonic effects}
\begin{figure}[H]
	\centering
	\subfloat{(a)}
	{\includegraphics[scale=0.275]{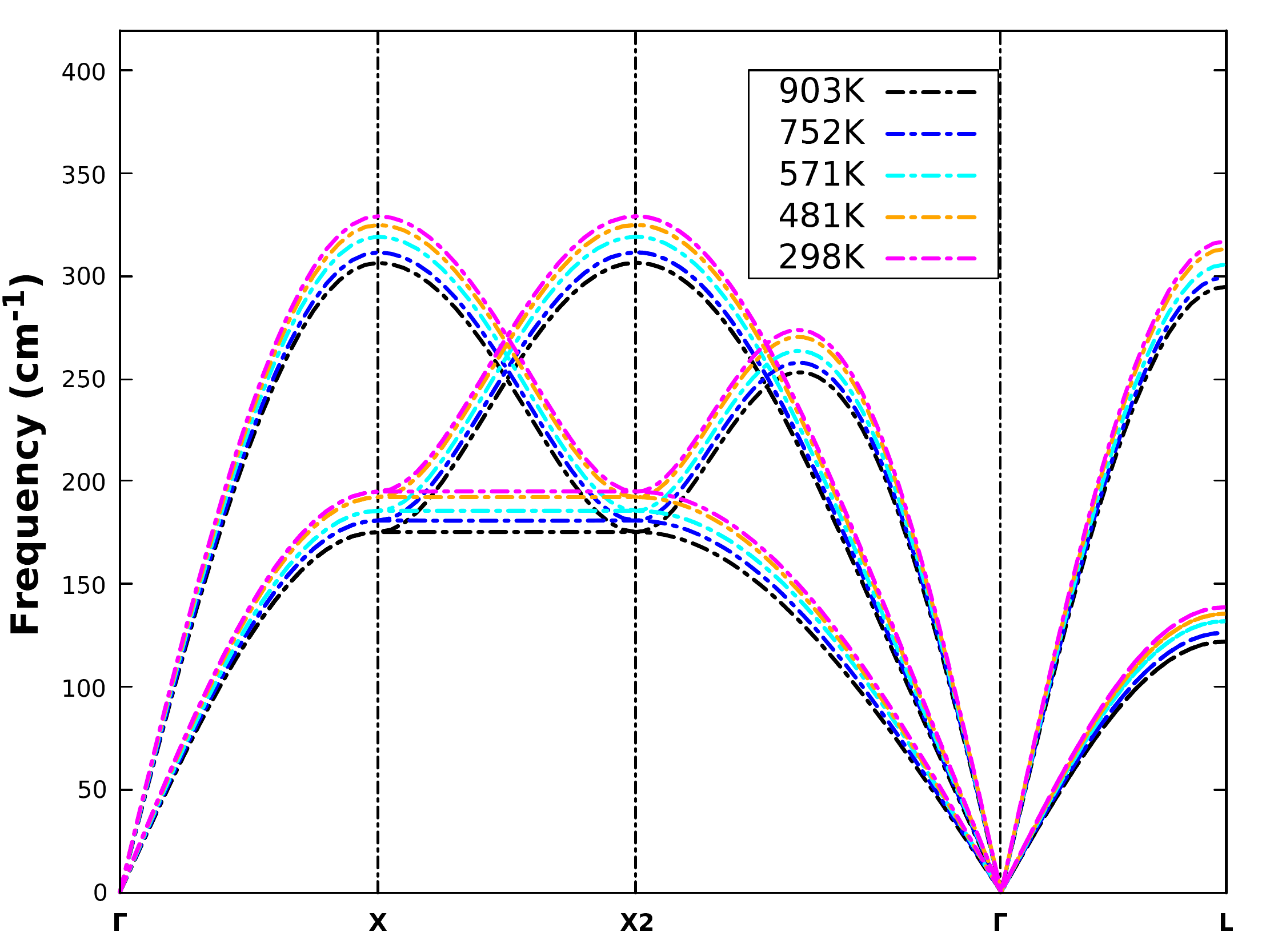}}
	\subfloat{(b)}
	{\includegraphics[scale=0.275]{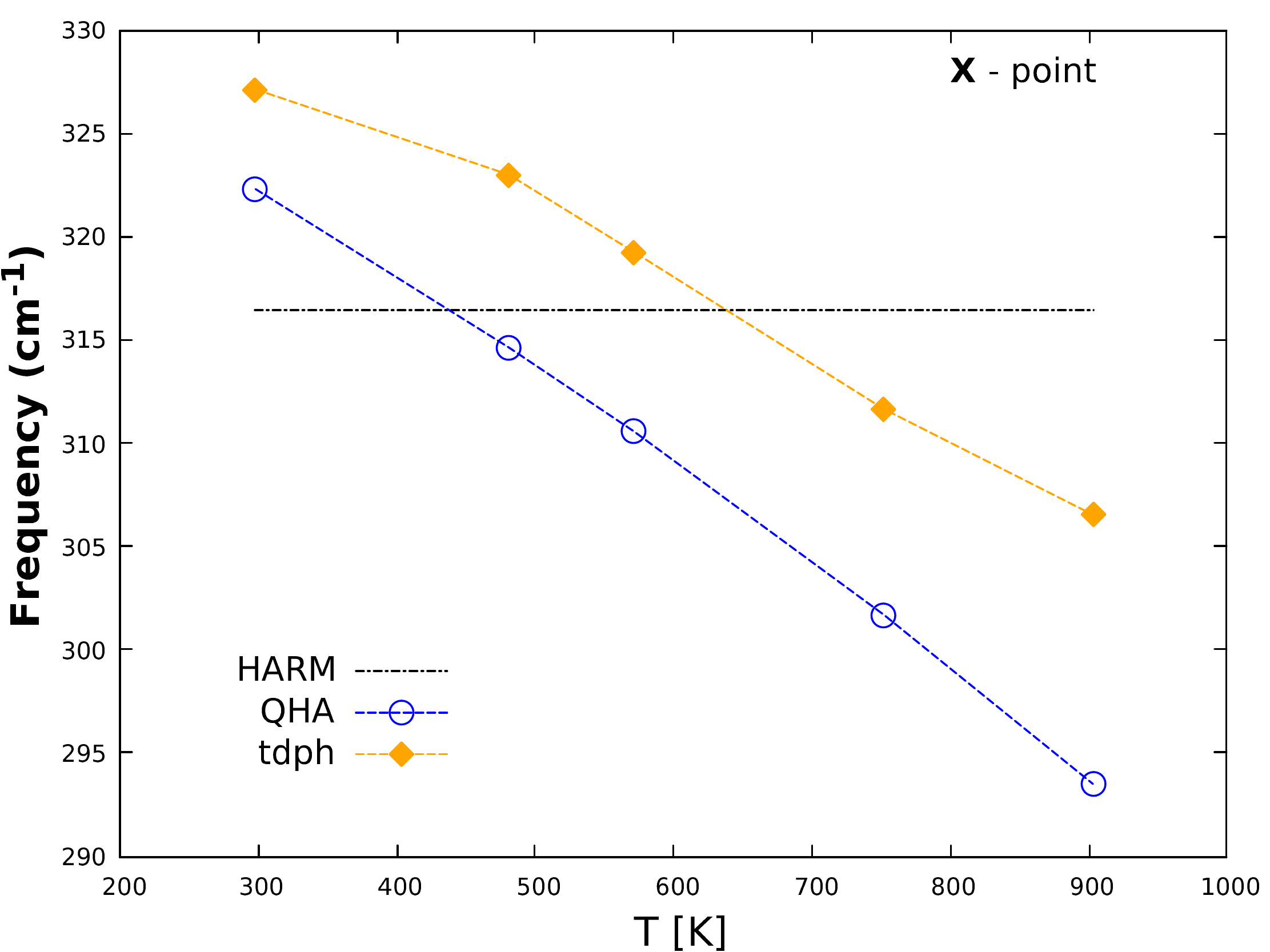}}
	\caption{The TDPH method gives the temperature dependence of phonons in Al, considering both anharmonic and quasiharmonic effects (using experimental lattice constants from  Ref.~\cite{nenno1960detection} (a) phonon dispersion (b) Evolution of phonons at finite temperature, \textbf{X}(0.5, 0.0, 0.5).}\label{fig:Al_shift}
\end{figure}

FCC aluminium exhibits weak anharmonicity. In this context, it is desirable to understand the degree to which quasiharmonicity and intrinsic anharmonicity affects phonon frequencies. Within the harmonic approximation, the effect of temperature is absent. Quasiharmonic approximation introduces temperature dependence via volume (thermal expansion) which tends to shift phonons as temperature is increased.
In the case of aluminum, the phonons become softer at higher temperature as observed in experiment~\cite{fultz2008phonons}. 
\\
To what degree do the anharmonic phonon frequencies change from their harmonic and quasiharmonic values at higher temperature? To answer this we apply QHA and TDPH method to aluminum at five different volumes corresponding to the temperatures of 273 K, 293 K, 571 K, 752 K and 903 K, obtained from the experimental data of Ref.~\cite{nenno1960detection} (See Table \ref{table:alat_T} and Fig. \ref{fig:Al_shift}).

\begin{figure}[H]
	\centering
	\subfloat{(a)}
	{\includegraphics[scale=0.385]{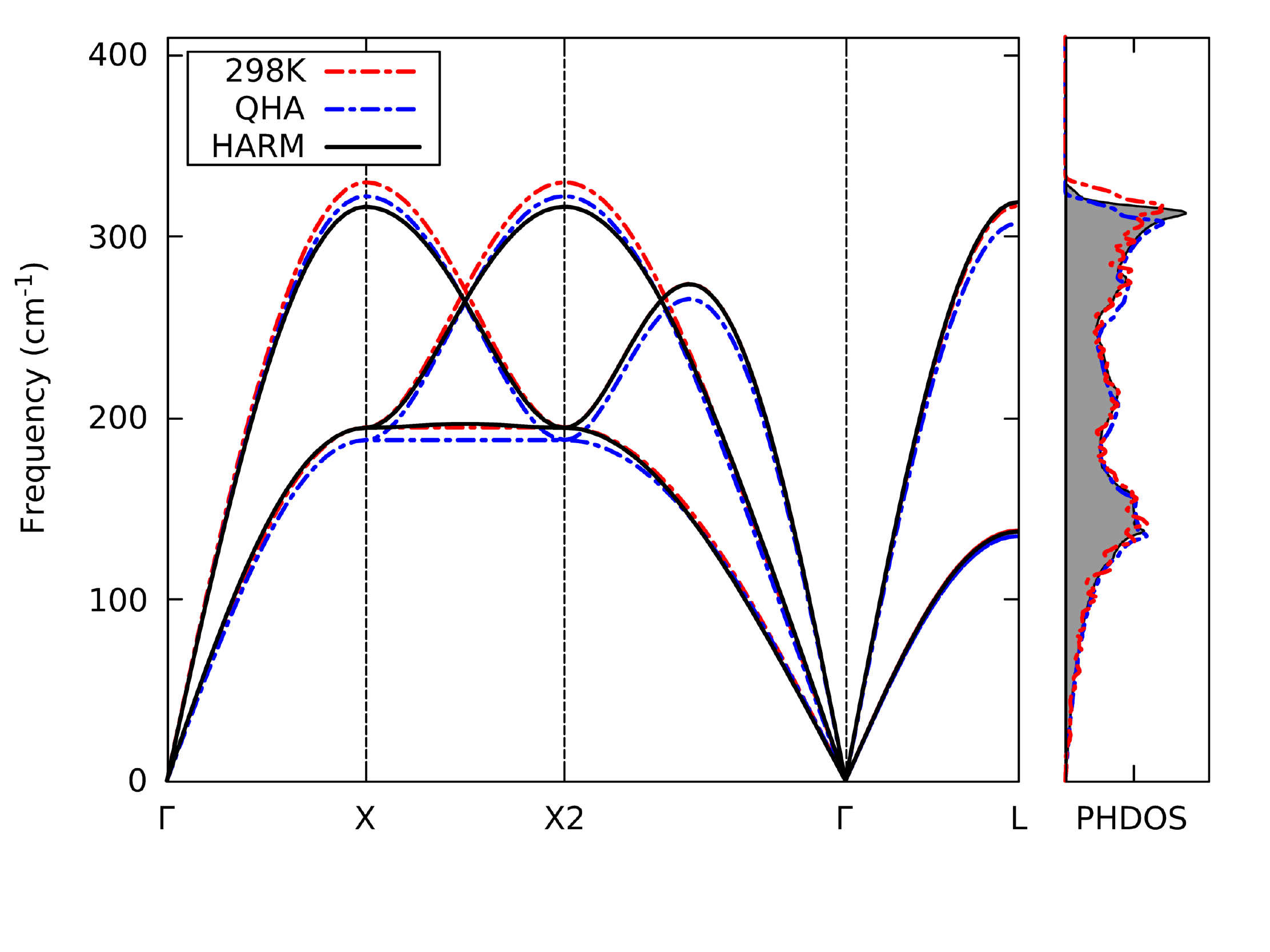}
		\label{fig:Al_298K}}
	\subfloat{(b)}
	{\includegraphics[scale=0.3854]{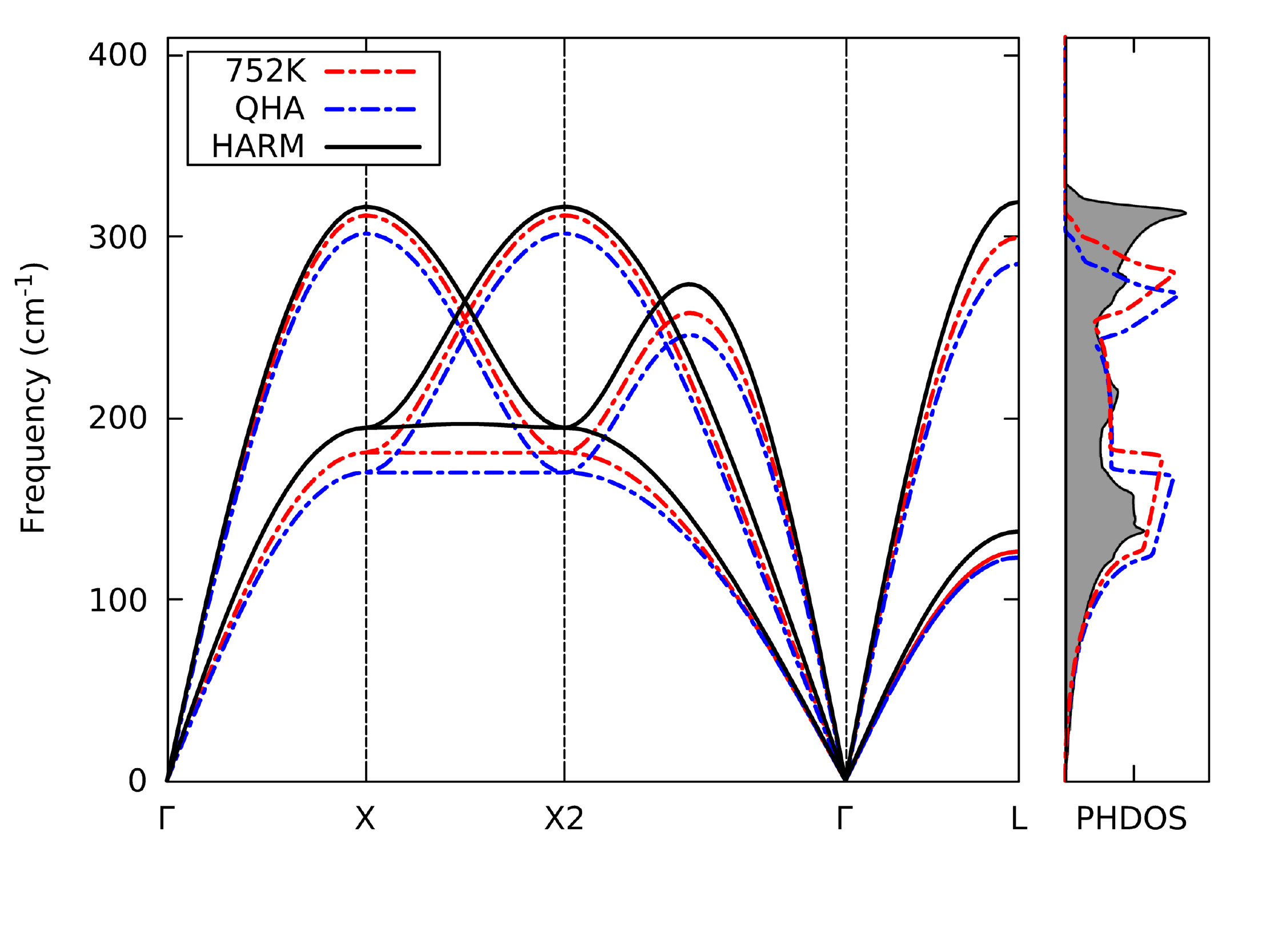}}
	\subfloat{(c)}
	{\includegraphics[scale=0.385]{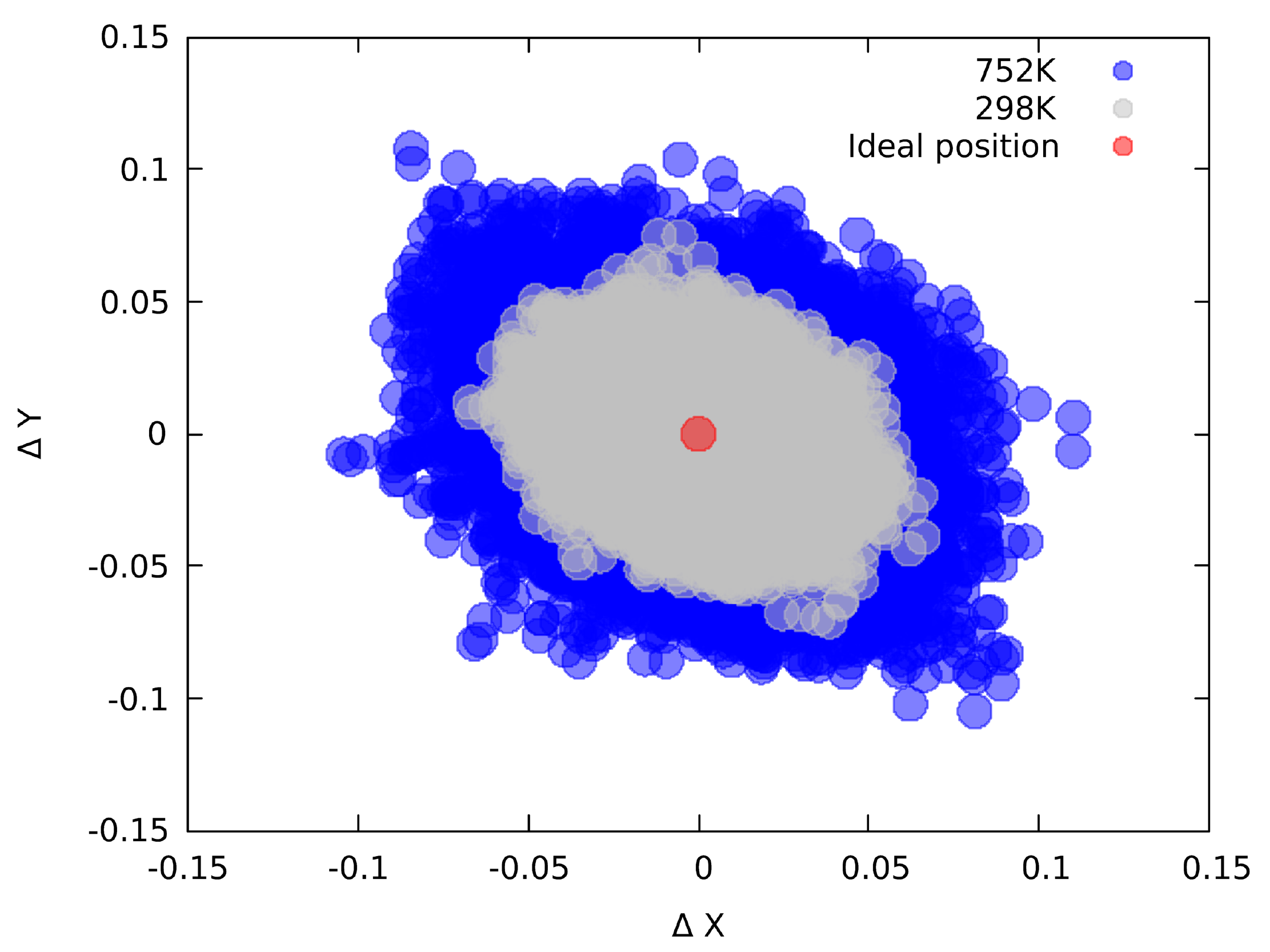}}
	\caption{Harmonic (HARM), Quasiharmonic (QHA) and anharmonic (TDPH) phonon dispersion of Al at (a) 298 K (b) 775 K (c) A two-dimensional projection of the trajectory of a typical atomic displacement in a $2 \times 2 \times 2$ supercell during AIMD at 298 K and 775 K. }\label{fig:Al_298K_vs_775K}
\end{figure}

The phonon dispersions calculated at 298 K using QHA and TDPH agree fairly well with each other, as shown in Fig.~\ref{fig:Al_298K_vs_775K}. The small difference indicates that thermal expansion is the dominant effect over anharmonicity at that temperature (see Appendix B). This is expected, given that 298K is lower than the Debye temperature of aluminum of 433K (i.e $\approx 0.6 \theta_D$) and the atomic vibrations from equilibrium can be represented fairly well with harmonic potential. However, at 752K ($\approx 1.8 \theta_D$), there is clear separation of QHA from the anharmonic shift. Essentially, QHA phonons are softer, and TDPH gives a better prediction that lies between the harmonic and QHA results. We note here that the TDPH result contains both quasiharmonic and  anharmonic corrections: 
\begin{equation}\label{eq:omega_anh}
\omega_{\texttt{tdph}}(V,T) = \omega_{\texttt{har}}(V) + \Delta\omega_{\texttt{qha}}(V(T)) + \Delta\omega_{\texttt{anh}}(V,T). 
\end{equation}
%
\begin{center}
	\captionof{table}{Comparison of the temperature dependence of lattice constant in Al using QHA, TDPH and experimental results from Ref.~\cite{nenno1960detection}. $a_{TDPH}$ is obtained from the minimum of the QHA free energy with renormalized phonons from TDPH.}
	\begin{tabular}{ m{2cm} m{2cm}  m{2cm} m{2cm} }
		T(K)   & $a_{EXP}$ & $a_{QHA}$   & $a_{TDPH}$\\ \hline \hline
	298.15 & 4.04962 & 4.07282  & 4.06131 \\
        481.15 & 4.06801 &	4.09353 & 4.07637 \\
        571.15 & 4.07788 &	4.10484 & 4.08338 \\
        752.15 & 4.09991 &	4.12999 & 4.09839 \\
        903.15 & 4.12039 &	4.15388 & 4.11210 \\
		\hline
 	\end{tabular}\label{table:alat_T}
\end{center}
\subsection{Zirconium: High-temperature $\beta$-phase }
Ti, Zr, and Hf have hexagonal closed-packed (hcp) crystal structure at ambient conditions which transform to body-centered cubic (bcc) structure at 1155 K, 1136 K, and 2030 K, respectively. \emph{Ab initio} calculation of their phonon spectra in bcc structure shows dynamical instability within (quasi)harmonic approximation~\cite{ozolins2012}.
When compressed at room temperature, hcp Zr ($\alpha$-phase) transformed to hexagonal ($\omega$-phase) at 17~GPa and subsequently to bcc ($\beta$-phase) above 35 GPa. However, there is a temperature-induced $\alpha \rightarrow \beta$ phase transition around 1136 K at 0 GPa~\cite{Bouchet2020betaZr}. 
\begin{figure}[H]
	\centering
	{\includegraphics[height=3in,width=4in]{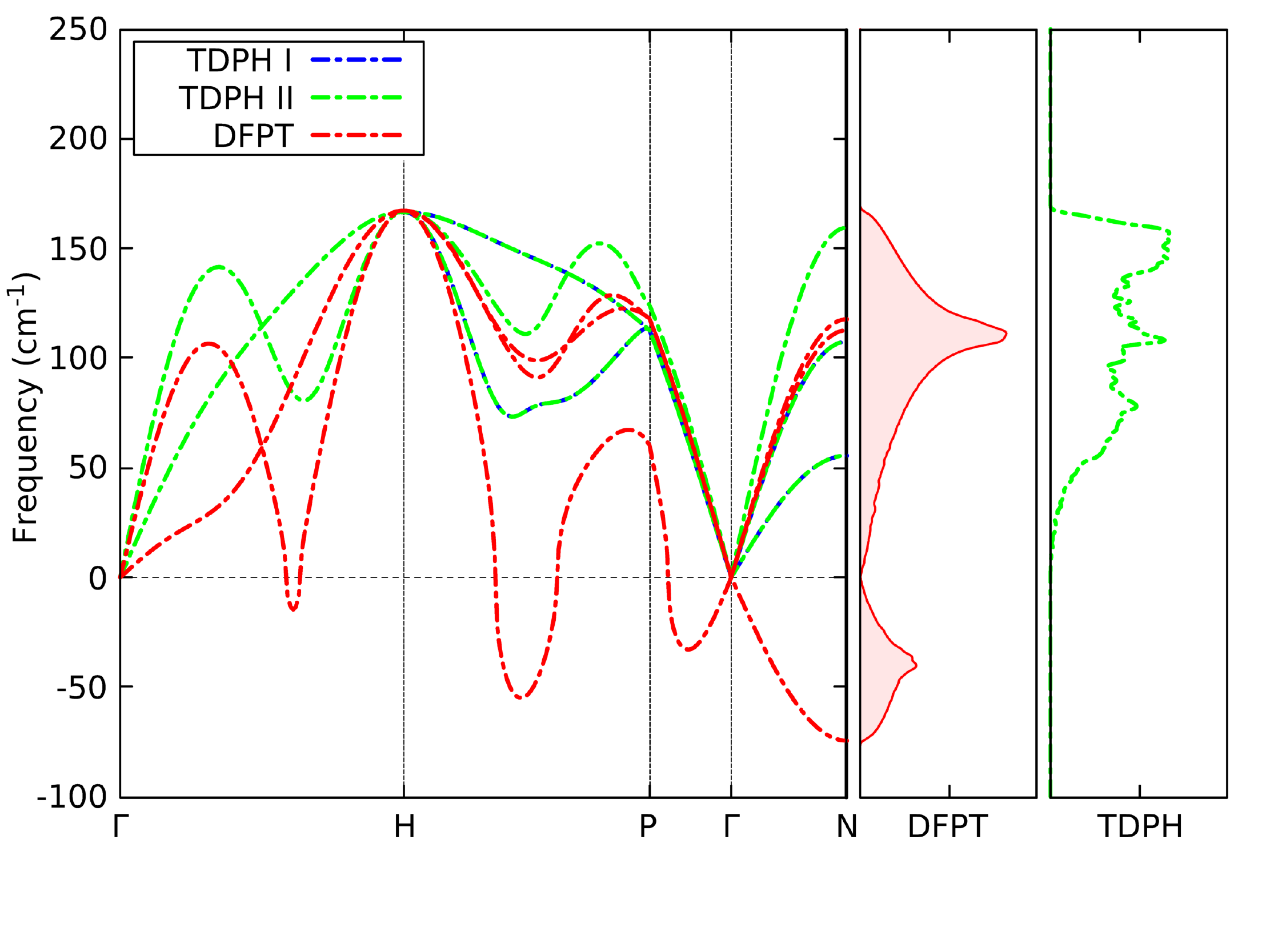}}
	\caption{ TDPH method applied to dynamically unstable bcc phase of Zr at 1200K, using initial force constants from DFPT (blue dash lines) and force-force correlators  (green dashed lines) from Ref.~\cite{pioud_2021}. Orange dashed lines represent harmonic phonons from DFPT. }\label{fig:gr-zr1200K}
\end{figure}
The high-temperature $\beta$-Zr is entropically stabilized and therefore strongly anharmonic.
Here, we will test the validity of TDPH by examining the role of anharmonicity in stabilizing the experimentally observed high temperature bcc phase of Zr. Figure \ref{fig:gr-zr1200K} shows the harmonic (orange dashed lines) indicating dynamical instability due to imaginary frequencies, and anharmonic phonon spectra and DOS. The TDPH method yields renormalized phonons at 1200 K.
Figure \ref{fig:gr-zr1200K} also shows how the TDPH works with two different initial guesses of the FCs: DFPT (blue-dashed lines) and the force-force correlator method of Ref.~\cite{pioud_2021} (green-dashed lines) represented as $\texttt{TDPH I}$ and $\texttt{TDPH II}$, respectively. A comparison of phonons using TDPH  vs force-force correlators method for Al and SrTiO$_3$ is given in Appendix C. Irrespective of the initial guess of the FCs, the TDPH method always gives converged results, consistent with the temperature at which the PES is sampled. 
\subsection{SrTiO$_3$: Renormalized Phonons in c-SrTiO$_3$ }
\begin{figure}[H]
      \centering
      {\includegraphics[height=3in,width=4in]{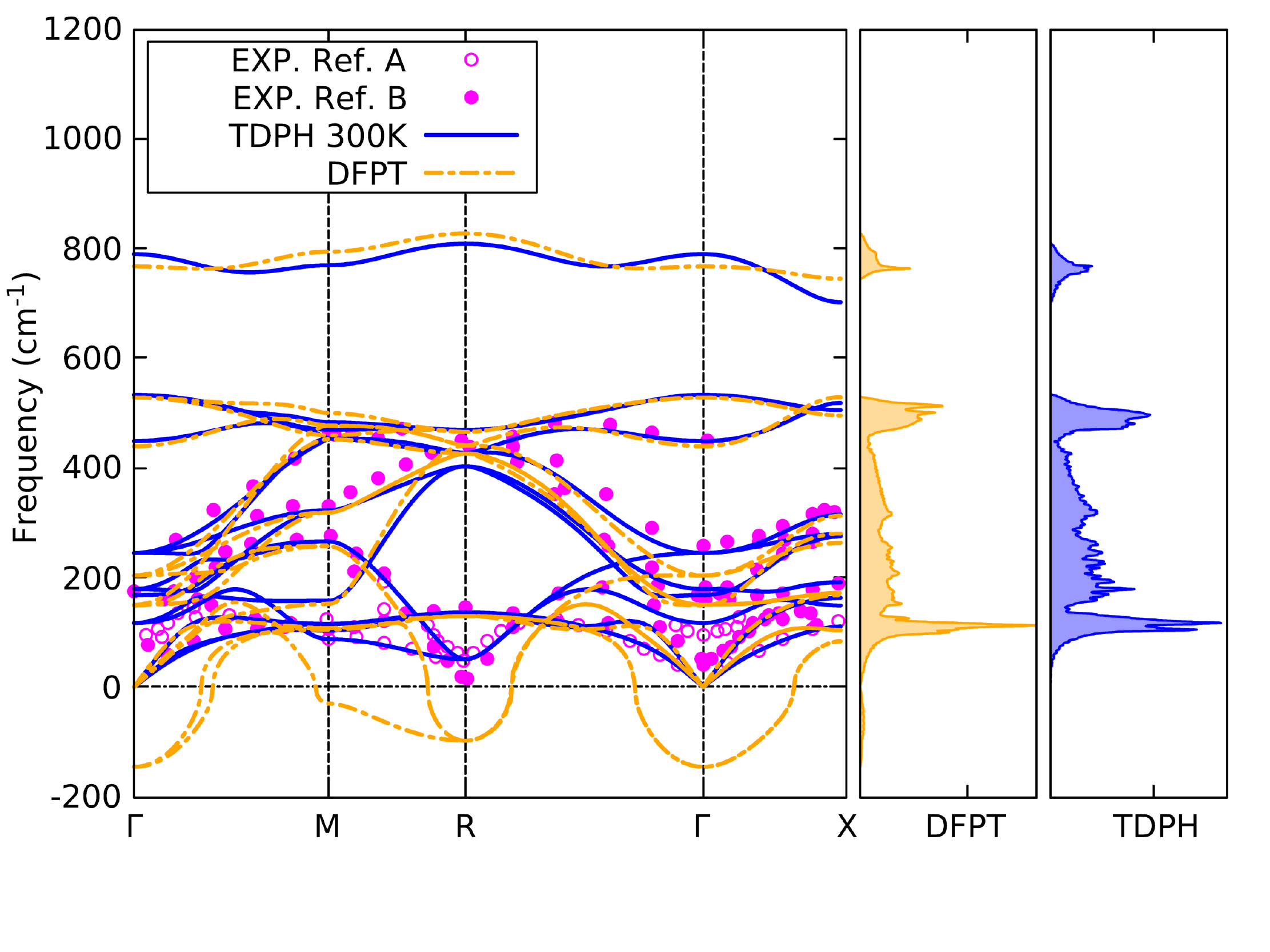}}
	\caption{Phonon dispersion and DOS of cubic phase of SrTiO$_3$. The dotted orange lines show the results based on the DFPT, and the solid lines represent the finite-temperature phonons obtained with the TDPH scheme at 300 K, consistent with experimental INS results of Ref.~A~\cite{stirling1972sto_INS_297K} and Ref.~B~\cite{cowley1969sto_110K}.}\label{fig:sto_exp}
\end{figure}

As a final benchmark of the TDPH method, we study the role of anharmonicity in the vibrational and thermal transport properties of SrTiO$_3$. Because of its various applications and rich physics, SrTiO$_3$ have been widely studied both theoretically and experimentally. At room temperature, SrTiO$_3$ 
has a cubic structure (c-STO), which transforms to low symmetry tetragonal phase (t-STO), below $T_c=105$K. First-principles method based on conventional harmonic and quasiharmonic approximation shows that c-STO is dynamically unstable, with soft modes at \textbf{$\Gamma$}, \textbf{R} and \textbf{M}-points as shown in Figure \ref{fig:sto_exp}. The antiferrodistortive (AFD) \textbf{R}-mode is associated with symmetry lowering cubic-to-tetragonal structural phase transition, while the  $\mathbf{\Gamma}$-mode is ferreoelecreic (FE). Previous studies indicated that including anharmonic effects can renormalize phonons, predict the cubic-tetragonal $T_c$~\cite{Tadano_2015}, enhance accurate prediction of carrier mobility~\cite{TDEP2018long-range_sto}, thermal conductivity~\cite{chen2021sto_kl} and band-gap dependence on temperature~\cite{tadano_wissam2020sto_Eg}.

Here, we applied the TDPH method to compute renormalized phonons in c-STO and lattice thermal conductivity with finite-temperature FCs.
In Figure \ref{fig:sto_exp}, we show the dynamic stabilization of the cubic phase, as indicated by renormalized positive frequencies relative to imaginary modes obtained from harmonic approximation using DFPT.

\begin{figure}[H]
	\centering
	\subfloat{(a)}
	{\includegraphics[scale=0.275]{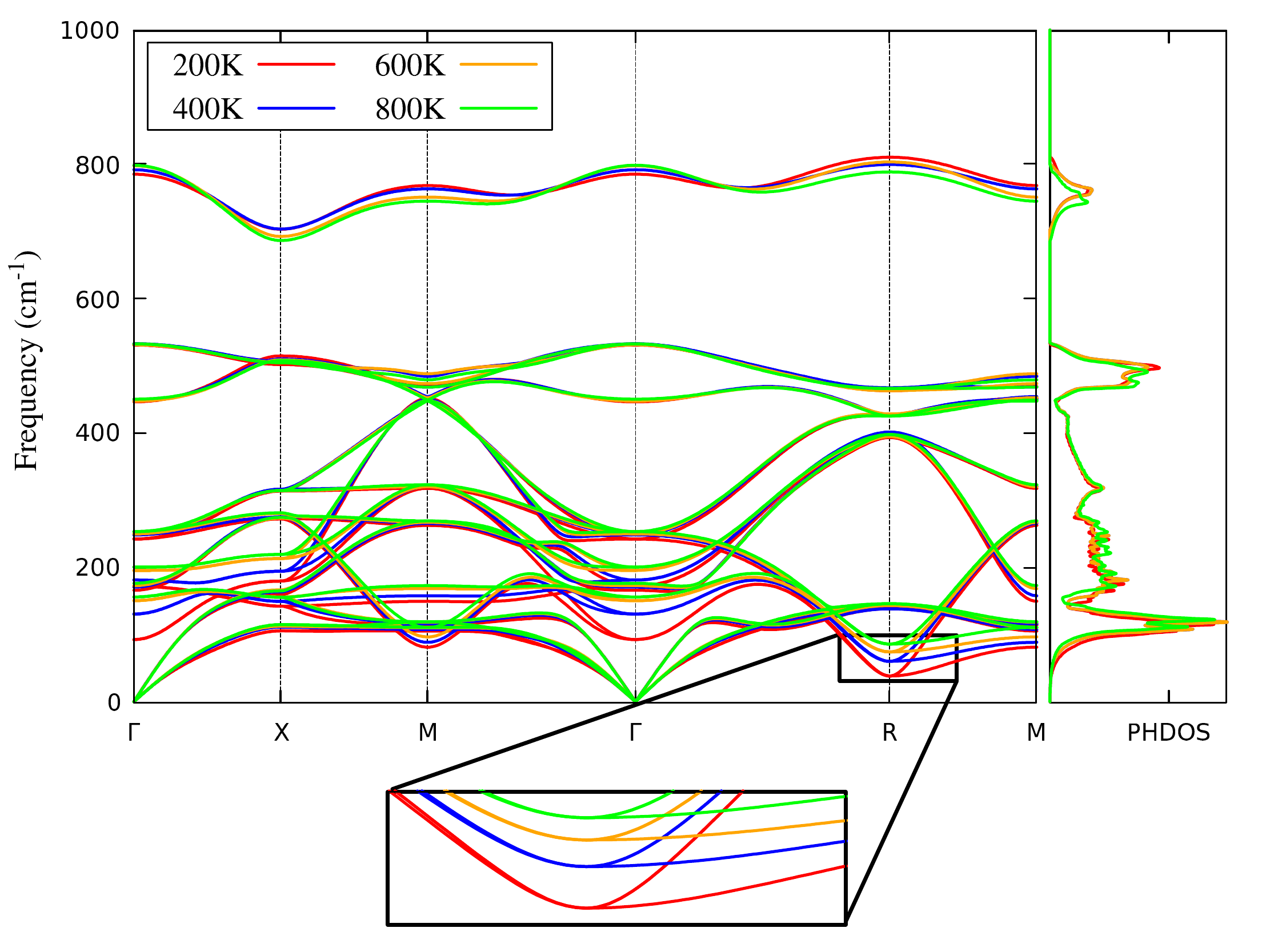}}
	\subfloat{(b)}
	{\includegraphics[scale=0.275]{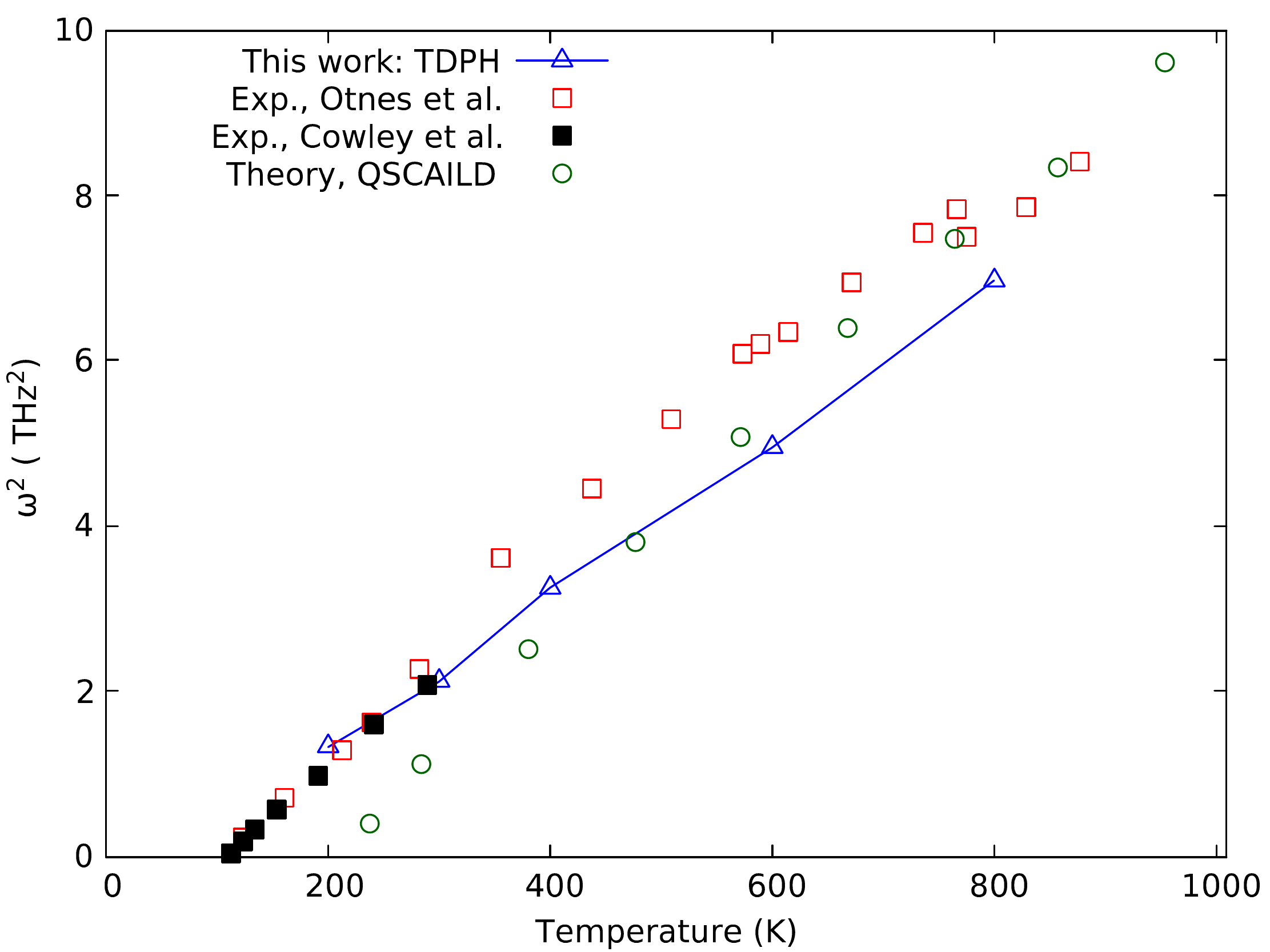}}
	\caption{(a) Temperature-dependence of phonon dispersion and DOS of cubic SrTiO$_3$ using TDPH method and a zoomed-in region around \textbf{R}-point. \\(b) Temperature-dependence of the squared frequency of the soft \textbf{R}-mode, compared with experimental and QSCAILD results.} \label{fig:sto_T_cw}
\end{figure}

The temperature-dependence of the phonon dispersion and density of states (DOS) in cubic SrTiO$_3$ is shown in Figure \ref{fig:sto_T_cw}(a), indicating phonon hardening with increasing temperature 
from 200 K to 800 K.
This effect has also been observed in previous experimental and theoretical studies~\cite{delaire2020anharmonic_sto}.

The temperature dependence of the squared frequency of the R mode has been used to predict the cubic-to-tetragonal phase transition in SrTiO$_3$. Figure \ref{fig:sto_T_cw}(b) compares TDPH results to experimental measurements and other theoretical methods. Our transition temperature of 75 K (Figure \ref{fig:sto_T_cw}(b)) is closer to the experimental value of 105 K than SCPH (220 K)~\cite{Tadano_2015} and QSCAILD (200 K)~\cite{QSCAILD_II_2021mingo} methods. The difference could be due to the functional used, as the lattice dynamical properties of ferroelectric materials are particularly sensitive to the DFT functionals~\cite{Vanderbilt_1995BaTiO3, Kresse2008SrTiO3}.
A recent attempt to circumvent the functional dependence of the FE and AFD instabilities in SrTiO$_3$ employed the SSCHA method and machine learning force field trained on random-phase approximation (RPA), obtaining a transition temperature of 172 K via Curie-Weiss fit of the AFD mode~\cite{verdi_2023}. 

\begin{figure}[H]
	\centering
	{\includegraphics[height=3in,width=4in]{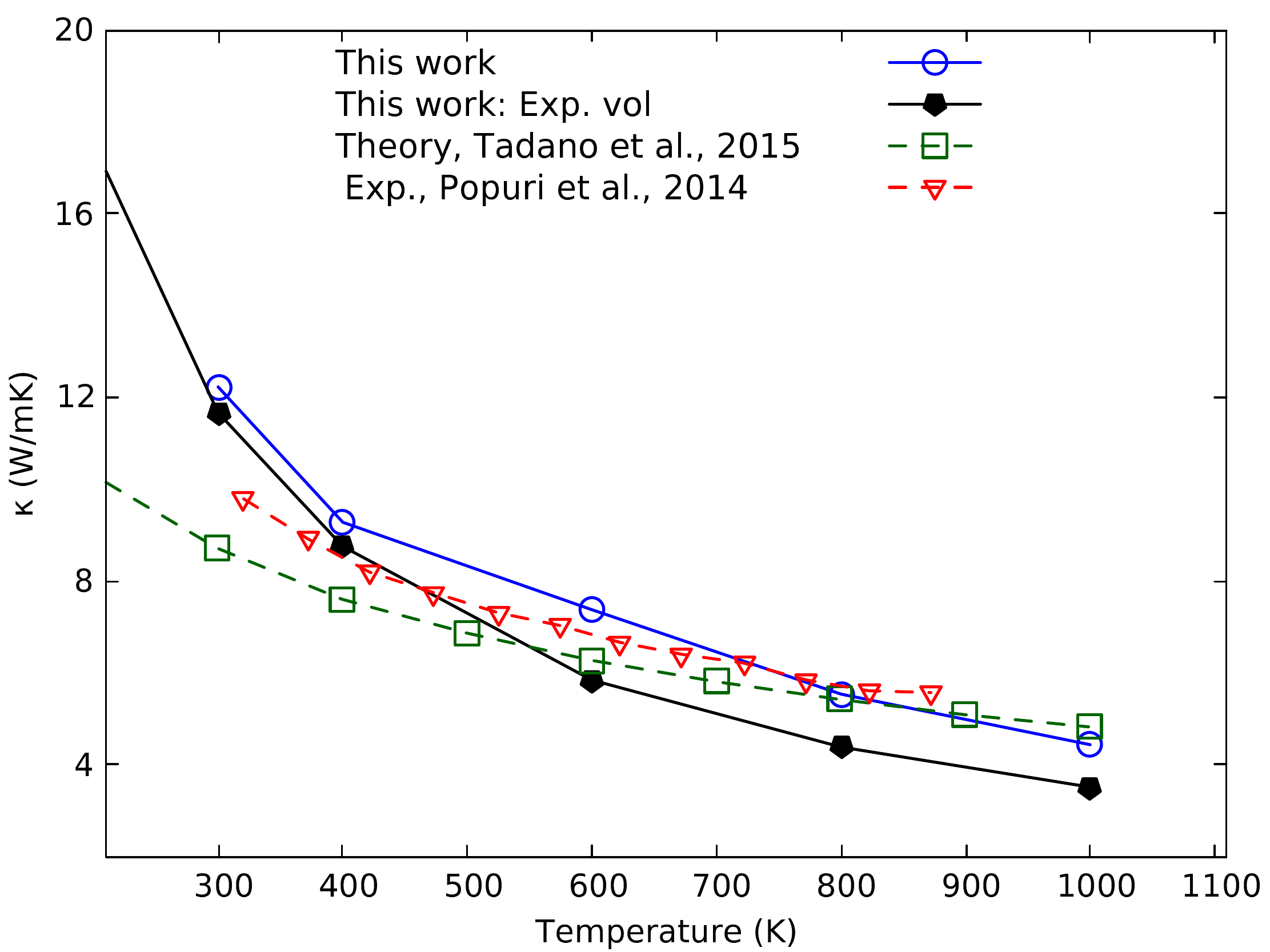}}
	\caption{Lattice thermal conductivity of cubic SrTiO$_3$ computed from BTE using temperature-dependent 2nd-order FCs and 3rd-order from perturbation theory and $(2n+1)$ theorem (See text for details). Red triangles are experimental data from Popuri \textit{\emph{et al.}},~\cite{kl_sto_popuri2014} and dark-green rectangles are from SCP theory\cite{Tadano_2015}. Blue circles indicate results at equilibrium DFT volume while black pentagon is at experimental volume.}\label{fig:sto_kl}
\end{figure}

We also compute the lattice thermal conductivity of SrTiO$_3$ based on the Boltzmann transport equation (BTE) of phonons in the single mode approximation (SMA) using renormalized 2nd-order FCs from TDPH, and compared to results from SCP theory~\cite{Tadano_2015} and experiment~\cite{kl_sto_popuri2014}.The effect of four-phonon scattering, which has been recently reported to suppress thermal conductivity in SrTO$_3$ by $15-20\%$ beyond 200 K~\cite{yue_chen2021srtio3_kl}, is not taken into account in our theoretical prediction. Nevertheless, using TDPH-based finite temperature FCs provides a reasonable estimate of thermal conductivity in SrTiO$_3$, which is impossible with harmonic FCs, as shown in Fig.~\ref{fig:sto_kl}.
\section{Conclusion}
We have developed an implementation of the TDEP method which uses a reciprocal-space minimization to reliably optimize the effective potential even for thousand of degrees of freedom and millions of data points with a negligible computational cost. This is achieved
by using the symmetry of the crystal to reduce the reciprocal-space \qq-point sampling of phonons to its irreducible wedge and, then for each \qq-point to reduce the dynamical matrix to its irreducible representations, expressed as a basis of symmetric hermitian ``dynamical'' matrices. The minimization itself is performed using the numerical gradient of the force residual with a new implementation of the Modified Levenberg-Marquardt algorithm which can be executed in parallel with ScaLAPACK.

The implementation has been tested on top of different dynamical sampling methods: uncontrolled NVE \emph{ab initio} molecular dynamics, NVT with a stochastic SVR thermostat, and Langevin Dynamics. 
We have shown that all three methods can produce universally valid temperature-dependent force constants. NVE is more eﬀicient, but with the strong caveat that it is susceptible to a large error from temperature drift. On the other hand,
by using the energy auto-correlation time to determine a suitable sampling interval between MD steps, the two controlled dynamic methods, i.e. NVT and LD, are similar in eﬀiciency, with the choice of the time step and thermostat parameters playing a crucial role in deciding if one can outperform the other.
In our test case, LD yielded the smoothest convergence for the phonon parameters as a function of the number of dynamics steps used for fitting.

We have tested the TDPH method in combination with Langevin dynamics, over a range of weakly and strongly anharmonic materials. We report that it can outperform QHA results in reproducing the temperature dependency of phonon frequencies in Aluminum. We can also correctly produce the high-temperature $\beta$ phase of Zirconium, with phonon frequencies which are virtually identical to those obtained from force-force correlation sampled over the same dynamics. The extrapolated frequency of the soft phonon band at the $R$ point predicts a phase transition temperature in good agreement with experiments and other simulation methods using comparable density functional approximations.

We recommend the combination of these two methods as an effective way to study finite-temperature phonon evolution, as they are readily accessible with only two computational parameters to tune, i.e. the time step and damping, for the reliability to optimize the force parameters and for their predictive power.

\section{Acknowledgements}
This work was granted access to the HPC resources of IDRIS under the allocation A0120907320 and A0100907320 made by GENCI. IBG fellowship is financed by the Petroleum Technology Development Fund (PTDF), Nigeria.

%
\appendix
\renewcommand{\thesection}{\Alph{section}}
\section{Computational Details}
\subsection{DFT and Phonon calculation}
DFT calculations were performed using the $\QE$ package~\cite{QE_2009, QE_2017}, employing optimized norm-conserving Vanderbilt pseudopotentials (NCPP)~\cite{pbe1996, ncpp2013Hamann} and ultrasoft pseudopotentials (USPP)~\cite{vanderbilt1990soft}. Electronic exchange and correlation are approximated with PBE~\cite{pbe1996} in Al and Zr, and PBEsol~\cite{pbesol2008} in SrTiO$_3$.
A cut-off energy of 100 Ry was employed for the wavefunctions in Al, Zr, SrTiO$_3$ whenever NCPP is used, while 30 Ry (Al), 50 Ry (Zr) and 75 Ry (SrTiO$_3$) are used in the case of USPP. Brillouin zone integration was performed using Monkhorst-Pack~\cite{Monkhorst-Pack_1976} k-point grid of $16\times 16\times 16$, $8\times 8\times 8$ and $4\times 4\times 4$ for the unit cells in Al, Zr and SrTiO$_3$, respectively. For supercell calculations, k-point grids were down-scaled proportionally. Given the metallic nature of Al and Zr, a cold smearing~\cite{marzari1999smearing} width of 0.05Ry was used.\\

The harmonic phonon spectra and FCs were computed using DFPT~\cite{DFPT_2001}, including long-range contributions to the dynamical matrices for polar SrTiO$_3$. In all cases, phonon $\qq$-grid were commensurate with supercell size, and strict  convergence was employed for phonon self-consistency $(10^{-16} Ry)$. In the case of cubic SrTiO$_3$, the FCs computed on $2 \times 2 \times 2$  $\qq$-grid were interpolated on $4 \times 4 \times 4$ grid for the computation of lattice thermal conductivity.
\subsection{Phonon lifetime and linewidth calculation}
The phonon linewidth and lattice thermal conductivity were computed on a $20 \times 20 \times 20$ $\qq$-mesh using \emph{Anharmonic} code of $\QE$~\cite{paulatto2013D3Q,paulatto2013k_l}. The lattice thermal conductivity of SrTiO$_3$ was computed using Boltzmann transport equation (BTE) within the Single-mode relaxation time (SMA) approximation along Cartesian direction $\alpha$ as
\begin{equation}
\kappa_l^{\alpha} = \frac{\hbar^2}{N_0\Omega\kappa_BT^2}\sum_{\gamma}\nu_{\alpha, \gamma}^2\omega_{\gamma}^2n(n+1)\tau_{\gamma}
\end{equation}
where $\hbar, N_0, \Omega, \kappa_B, T, n$ represent Planck constant, total number of $\qq$-mesh points, unit cell volume, Boltzmann constant, temperature, and Bose-Einstein phonon population. The phonon energy $\omega_{\gamma}$, and group velocity  $\nu_{\alpha, \gamma}$ are computed from  renormalized 2nd-order FCs using TDPH method. 

\subsection{Molecular and Langevin Dynamics}
We perform \emph{ab initio} molecular dynamics as implemented in $\QE$~\cite{QE_2009,QE_2017} using the same parameters and settings employed for the harmonic phonon calculations (pseudopotential, cut-off energy for wavefunctions, etc). The size of the supercell is commensurate with the phonon $\qq$-grid, while k-point grids
were down-scaled proportionally. Temperature is controlled by stochastic velocity rescaling (svr) method~\cite{svr2007canonical} to ensure efficient canonical sampling. The length of the simulation and number of MD snapshots depends strictly on the material. However, after equilibration, 100 snapshots sampled over 2000 steps (equivalent to 2 ps) were sufficient to obtain converged results in both Al and Zr.

The second method is Langevin dynamics (LD) with Bussi and Parrinello algorithm~\cite{Bussi_Parinello_2007LD} as recently implemented in Ref.~\cite{pioud_2021}. LD integrates the equation of motion with deterministic (DFT), stochastic, and frictional forces at a given temperature, based on the Trotter factorization of Liouvillian operator.
The Langevin damping $\gamma$ (see text) had to be chosen so that the sampling efficiency is optimized. We found $\gamma= 0.00146 \; a.u.$, to be optimal. 
\section{Anharmonic correction to QHA free energy }
We compute the effect of anharmonicity on the thermal expansion of Al by computing the free energy using quasiharmonic approximation (QHA) and temperature dependent phonon method (TDPH). The TDPH free energy is essentially the QHA free energy with renormalized phonons from TDPH. This appproach has a difficulty because the volume-temeprature equation of state used (VT-EOS) for the TDPH simulation may not be equal to the final one. This problem could be solved self-consistently, but we observe that a single-shot correction, starting from the standard QHA VT-EOS can produce a significant correction while remaining fully \textit{ab initio}.

\begin{figure}[H]
		\centering
		{\includegraphics[scale=0.45]{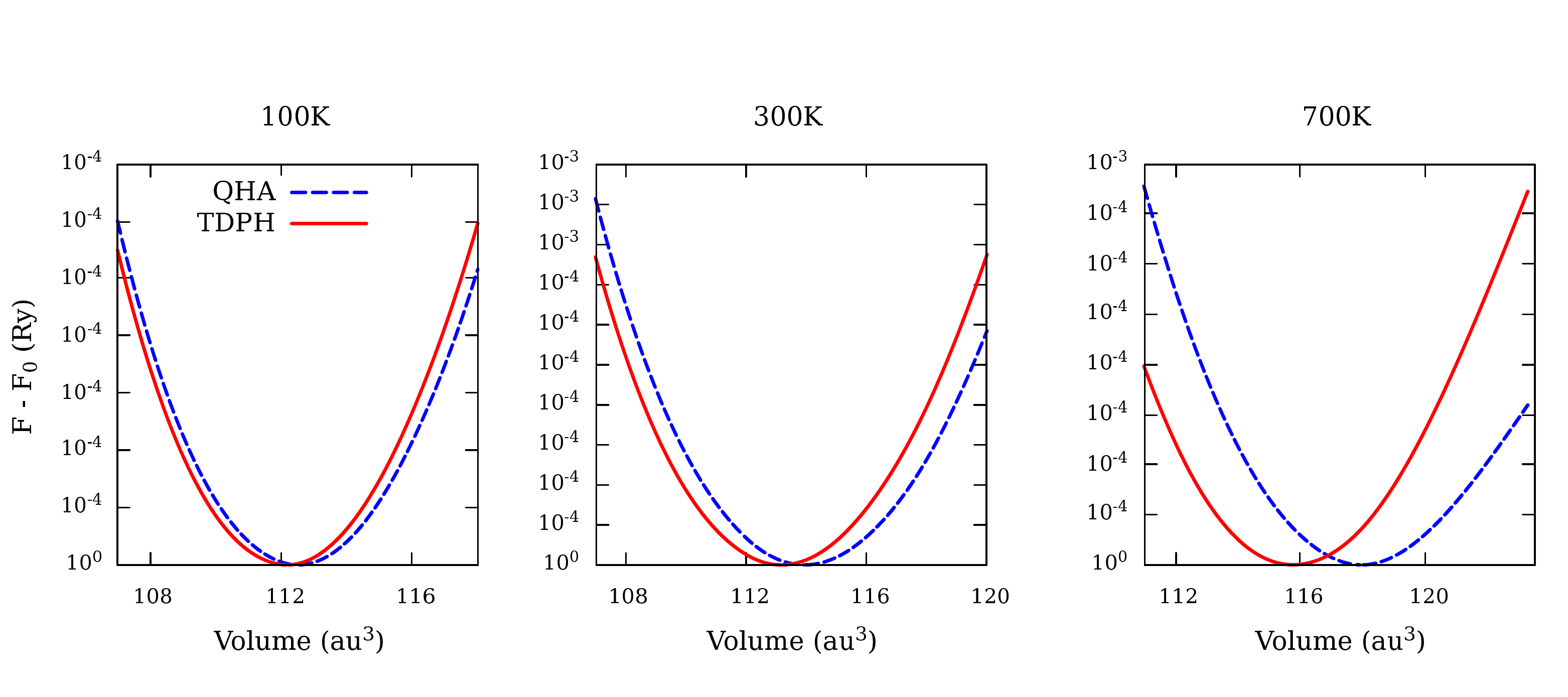}}
		\caption{The free energy of Al per unit cell vs lattice volume computed using quasiharmonic approximation (QHA) and temperature dependent phonon method (TDPH).  }\label{fig:gr-free_enegry}
\end{figure}

In Figure \ref{fig:gr-free_enegry} at low temperature (100 K and 300 K), the lattice volume corresponding to the minima of the free energy changes slightly, an indication of weak anharmonicity. At 700 K, the TDPH correction to the QHA equilibrium volume is more significant, 
indicating the importance of including anharmonic effects at higher temperature.
\section{Phonons from TDPH vs force-force correlators method}
We compare phonons from TDPH  with force-force correlators method~\cite{pioud_2021} in fcc Al and cubic SrTiO$_3$. Note that in the case of SrTiO$_3$ (Figure \ref{fig:TDPH_vs_ff}a ), the TDPH method is applied to short-range forces only, without the long-range effect due to dipole-dipole interaction which is removed from the initial guess of the harmonic force constant. This contribution is absent in the force-force correlators method as shown by the degenerate phonon bands at $\Gamma$-point. 
\begin{figure}[H]
	\centering
	\subfloat{(a)}
	{\includegraphics[scale=0.28]{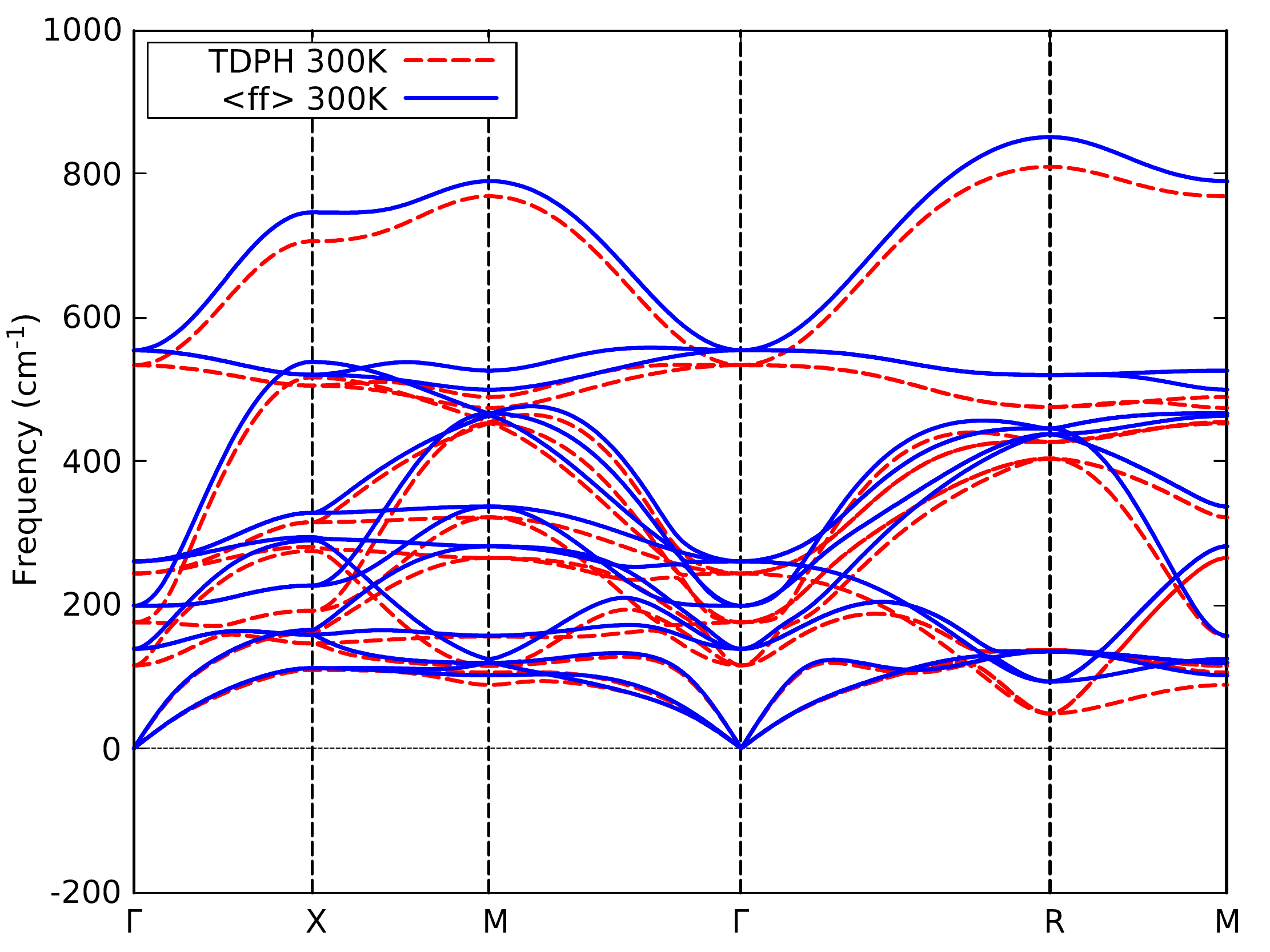}}
	\subfloat{(b)}
	{\includegraphics[scale=0.28]{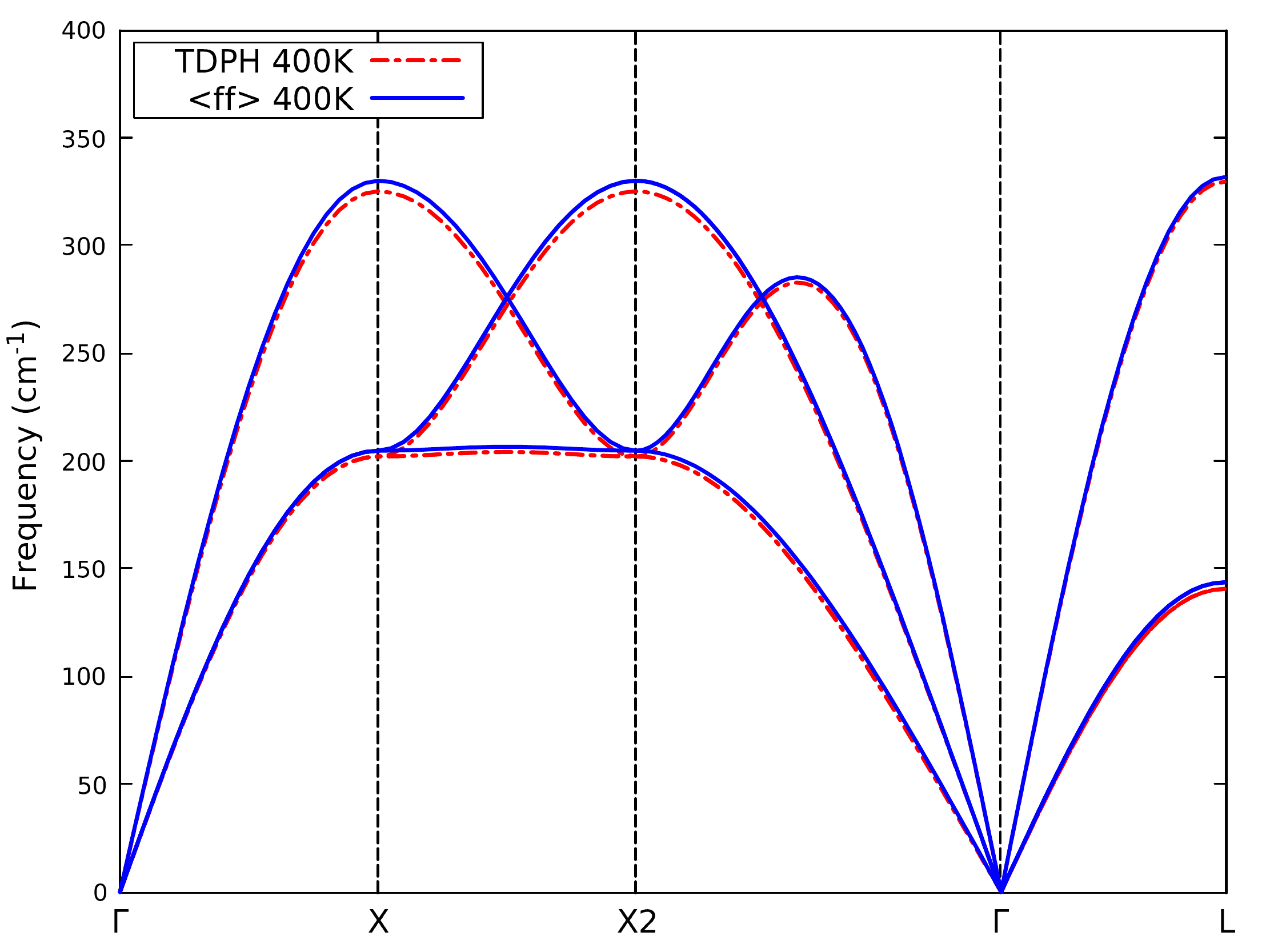}}
	\caption{Temperature-dependent phonon dispersion using TDPH and force-force correlators $<ff>$ methods~\cite{pioud_2021}. (a) cubic SrTiO$_3$ without long-range force-constants (b) fcc Al.} \label{fig:TDPH_vs_ff2}
\end{figure}
\section{Phonons and autocorrelation in the different dynamic simulations}\label{app:autocorr}
We compare the phonon dispersion in SrTiO$_3$ using three different methods of sampling the PES: LD, MD-NVT and MD-NVE. In each case, the 1,000 configurations are sampled using  $\tau_{\texttt{sampling}} = C(t,t')$, where $C(t,t')$ is the autocorrelation function computed using Wiener–Khinchin theorem
\begin{equation}
C(t,t') = \int_{-\infty}^{\infty}|E|^2 e^{-2\pi ivt} dv,
\end{equation}
where E is the instantaneous energy from \textit{ab initio} simulation. Large deviation of the average temperature of MD-NVE from the target (T = 300 K) in Table \ref{table:autocorr} is due to the absence of thermostat, and the small autocorrelation time comes from fast descend of NVE to minima (see text).
\begin{center}
\captionof{table}{Autocorrelation time and average Temperature}
	\begin{tabular}{ m{2cm} m{2cm}  m{2cm} m{2cm} }
		& LD & MD-NVT & MD-NVE\\ \hline \hline
  		$T_{avr}$(K) &    299 &	298 & 315 \\
		C(t,t') &    30 &	12 & 8 \\
		\hline
\end{tabular}\label{table:autocorr}
\end{center}
\begin{figure}[H]
		\centering
		{\includegraphics[scale=0.45]{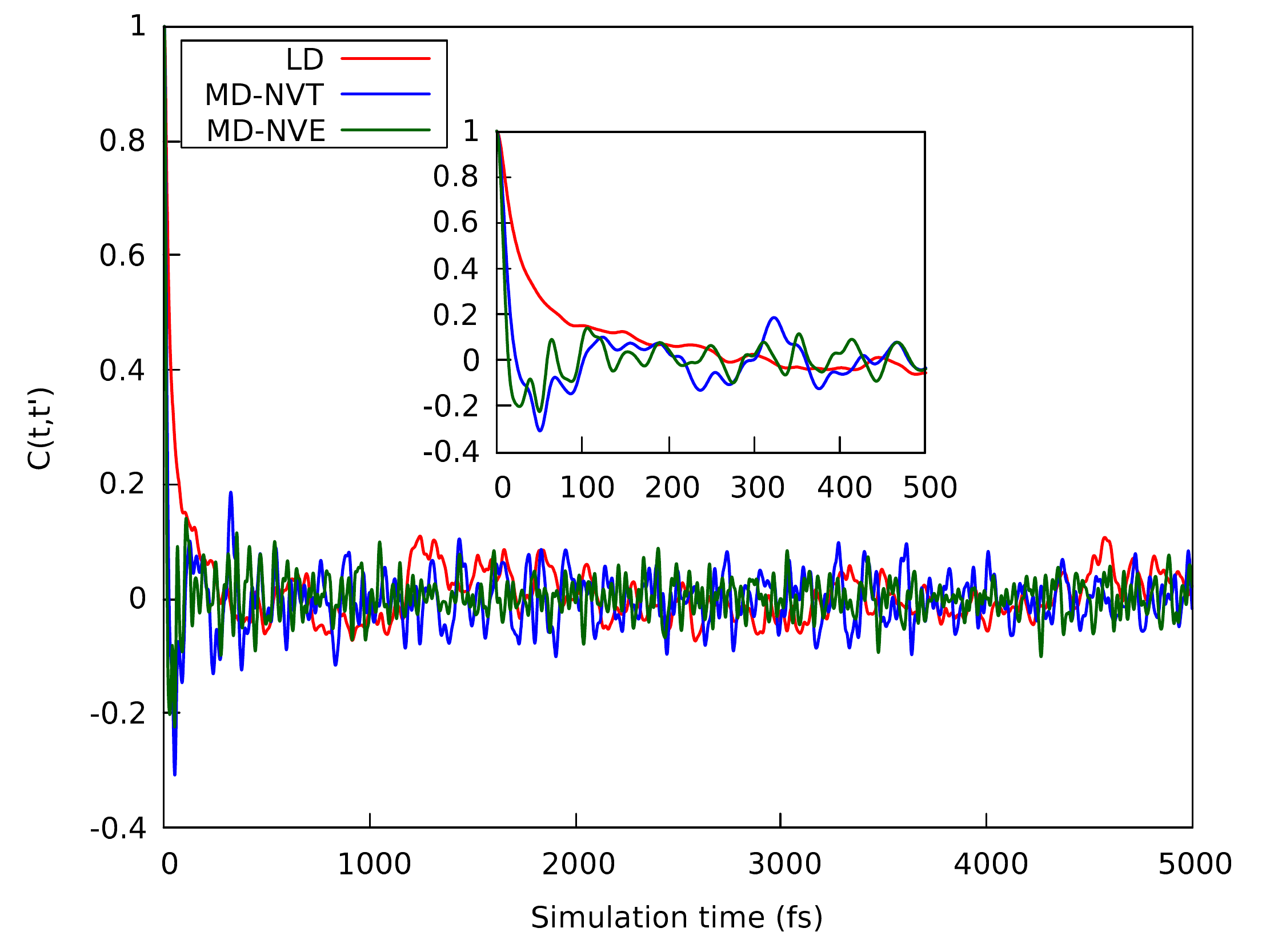}}
		\caption{Comparison of the autocorrelation function of the \textit{ab initio} energy from the three different dynamic simulations in cubic SrTiO$_3$. }\label{fig:gr-autocorr_all}
\end{figure}
\begin{figure}[H]
		\centering
		{\includegraphics[scale=0.45]{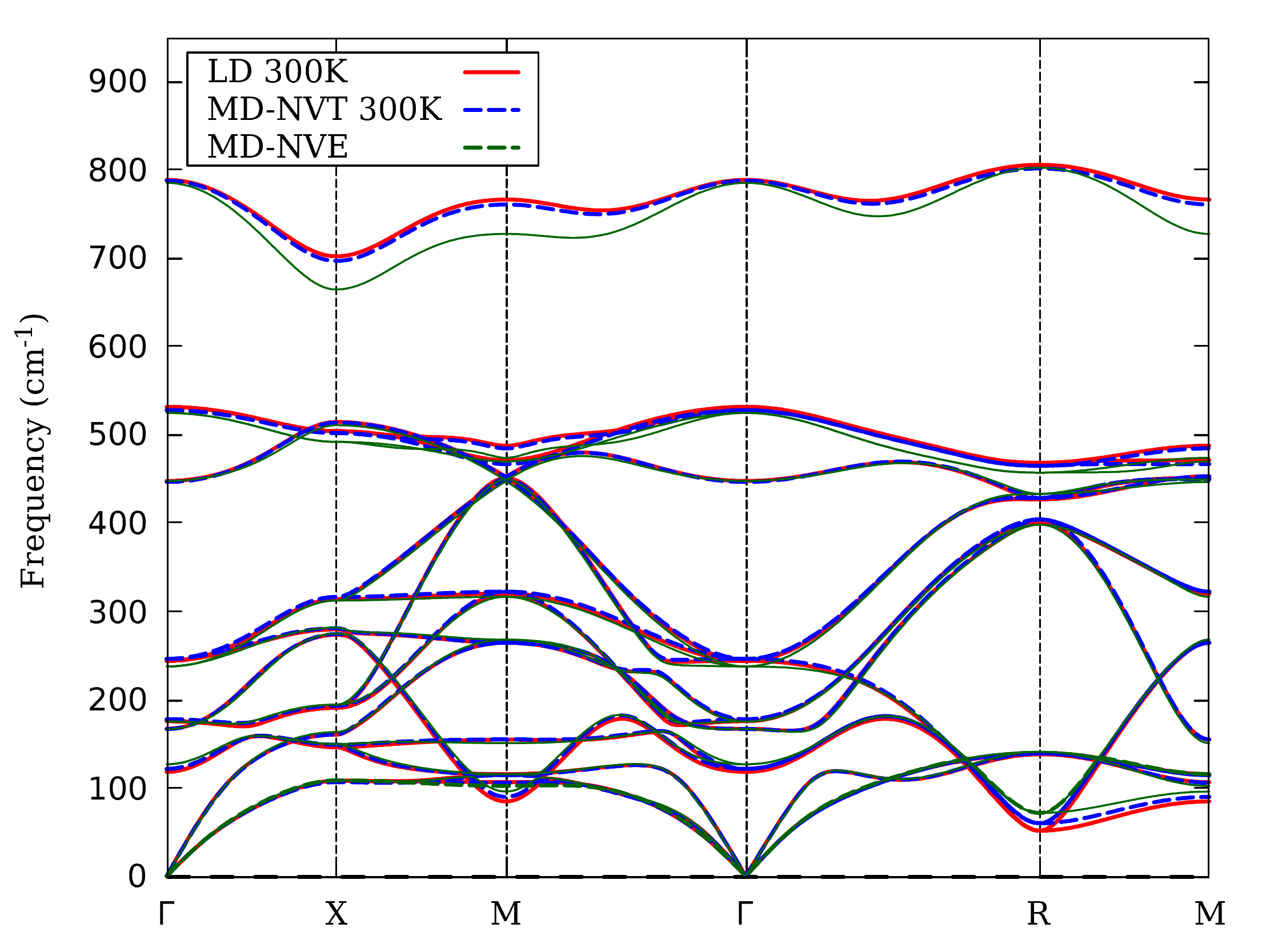}}
		\caption{Comparison of phonon dispersion from the three different dynamic simulations in cubic SrTiO$_3$.  }\label{fig:gr-all-compare}
\end{figure}
\section{TDPH input description}
This code reads a set of initial dynamical matrices for a given system
and optimizes the harmonic force constants over a series of images that
can be the output of a molecular dynamics calculation performed with $\QE$, or a Langevin Dynamics calculation from PIOUD
code. The code will expect that the size of the supercell of the dynamics simulation is the same as that of the force constants.
\begin{table}[H]
	\centering
	\caption[The first table]{TDPH input description}
	\setlength{\arrayrulewidth}{1pt}
	\begin{tabular}{m{2.6cm} m{12.cm}}
		\textbf{Input} &	\textbf{Description} \\
		\hline \hline
		\texttt{ai}   & \texttt{CHARACTER, "md" or "pioud"}: \texttt{"md"} if the sampling comes from \emph{ab initio} molecular dynamics using the standard $\QE$ dynamics engine (i.e. using calculation='md' in the pw.x input) or \texttt{"ld"} for Langevin dynamics based on the Bussi and Parrinello's algorithm engine. \\ 
		\hline
		\texttt{fmd}   &  \texttt{CHARACTER, "md.out"}: File from standard $\QE$ dynamics engine. \\ 
		\hline
		\texttt{ftau, fforce, ftoten}   & \texttt{CHARACTER, default "positions.dat" "forces.dat", "sigma.dat":} Output files from LD containing atomic positions in Angstrom, atomic forces in Hartree per Angstrom, and potential energy Hartree atomic units. \\
		\hline
		\texttt{file\_mat2} & \texttt{CHARACTER, default "mat2R":} File containing initial force constants. Must be periodic(i.e. generated with \texttt{nf = 0} using \texttt{d3\_q2r.x}).
		\\ \hline
		\texttt{nfirst, nskip, nmax}& \texttt{INTEGER, default 1, 100, 5000:} When reading MD or LD trajectory files, read one every \texttt{nskip} steps starting from \texttt{nfirst} until \texttt{nmax} configurations are read (i.e. from \texttt{nfirst} to \texttt{nfirst}+\texttt{nskip}$\times($\texttt{nmax}$-1)$).
		\\ \hline
		\texttt{fit\_type} & \texttt{CHARACTER, forces:} The fitting method described in the text. \\  
		\hline
		\texttt{minimization} & \texttt{CHARACTER, ph, ph+zstar or global:} These choices define the which TDPH minimization method to employ. \texttt{ph} will minimize short-range FCs and add long-range FCs if present (as they are). \texttt{ph+zstar} decompose both short and long-range FCs, minimize and subsequently recompose the TDPH FCs.  \texttt{global} is experimental.
		\\ \hline
		\texttt{e0}&   \texttt{REAL:} Equilibrium total energy from DFT before AIMD steps. \\ \hline
		\texttt{thr}   & \texttt{REAL, DEFAULT=1.d-12:} file from Quantum espresso containing initial atomic coordinates and  \\ \hline
	\hline
	\texttt{randomization}   & \texttt{REAL:} Adds or subtracts random numbers to initial phonon parameters.  \\ 
		\hline
	\end{tabular}
	\label{table:tdph_input}
\end{table}
\bibliography{./ref} 
\bibliographystyle{iopart-num}  
\end{document}